\documentclass[%
 reprint,           
 superscriptaddress,
 nofootinbib,       
 amsmath,amssymb,
 aps,
 pre,
]{revtex4-2}

\newcommand{\me}{m_{\rm e}}

\newcommand{\wpp}{\omega_{\rm p}}

\newcommand{\Ma}{M_{\rm A}}

\newcommand\redsout{\bgroup\markoverwith{\textcolor{red}{\rule[0.5ex]{2pt}{0.4pt}}}\ULon}
\newcommand{\entity}{\texttt{Entity}}

\usepackage{graphicx}
\usepackage{dcolumn}
\usepackage{bm}
\usepackage{xcolor}
\usepackage[T1]{fontenc}
\usepackage[utf8]{inputenc}
\usepackage{hyperref}
\usepackage{soul}
\usepackage{orcidlink}
\graphicspath{{./}{figures/}}

\AtBeginDocument{\def\selectlanguage#1{}}

\begin{document}

\title{Radiation Reaction in Relativistic Magnetized Shocks of Neutron Star Magnetospheres}

\author{Arno Vanthieghem\,\orcidlink{0000-0002-3643-9205}}
\email{arno.vanthieghem@obspm.fr}
\affiliation{Observatoire de Paris, Sorbonne Université, Université PSL, CNRS, LUX, F-75005 Paris, France}

\author{Amir Levinson\,\orcidlink{0000-0001-7572-4060}}
\affiliation{School of Physics and Astronomy, Tel Aviv University, Tel Aviv 69978, Israel}

\date{\today}

\begin{abstract}
The steepening of fast magnetosonic waves in the inner magnetospheres of magnetars can produce monster shocks, whose precursor emission may account for fast radio bursts (FRBs). For sufficiently bright bursts, particles crossing the shock radiate their energy within a single gyration period, raising the question of how radiation reaction affects the shock structure and its coherent emission. We investigate strongly cooled, relativistically magnetized shocks using a stationary multifluid model generalized to include the Landau--Lifshitz radiation-reaction force, together with particle-in-cell simulations of monster-shock dynamics. We show that the kinetic-scale shock transition is governed by only two dimensionless parameters: the Alfv\'enic Mach number and a cooling parameter $\tau$ that measures radiative losses over a gyration period. 
Coherent precursor emission is triggered above a critical Mach number that increases with $\tau$, at which the beam dynamics becomes chaotic.
Above this threshold, radiative cooling shifts the peak frequency of the precursor to higher values, scaling as $\tau^{2/7}$. Remarkably, monster shocks remain far above the emission threshold even under strong radiative cooling. Radiation reaction therefore does not quench the precursor emission; instead, it primarily shifts the spectral peak into the kinetic regime. The luminosity of the precursor wave is determined solely by the luminosity of the driving kHz wave and the conversion efficiency, $\epsilon_X\sim 10^{-3} - 10^{-2}$.
\end{abstract}


\maketitle

\section{introduction}

\begin{figure*}[t]
	\begin{center}
		\includegraphics[width=1\textwidth]{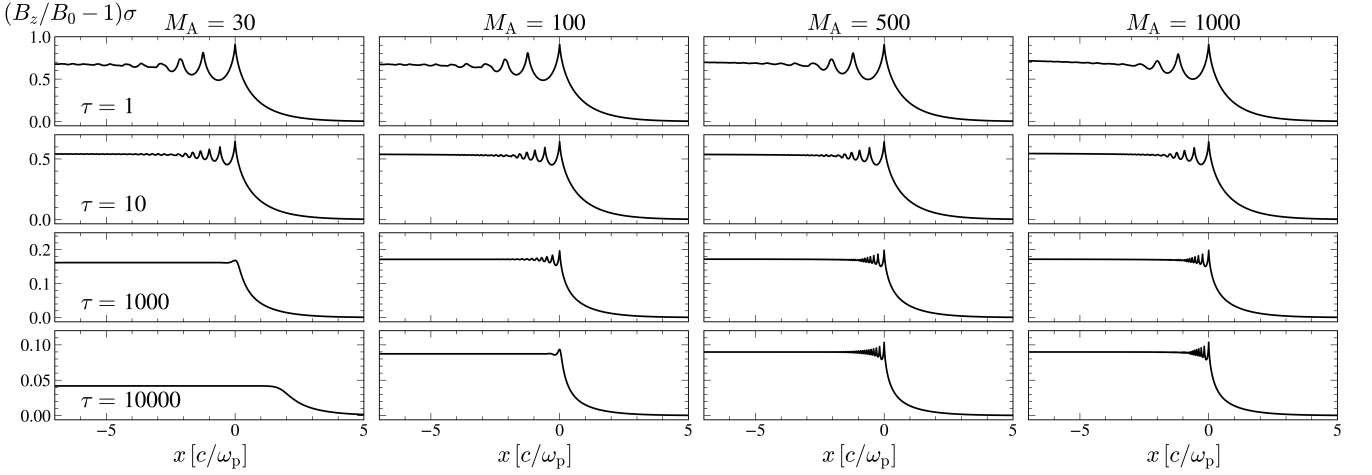}
		\caption{Multi-fluid stationary shock profiles for $N=10^3$ and $u_0=10^6$, shown for different values of the cooling parameter, $\tau=(\sigma_{\rm T}n_{\rm up}c/\omega_{\rm p})u_0\Ma^2$, and the Alfvénic Mach number, $\Ma$. At fixed $\tau$, the downstream magnetic compression, $B_z/B_0-1$, scales as $\sigma^{-1}$. The variation of the shock profile with $\Ma$ at $\tau \gg1$ illustrates the $\Ma$-dependence of the X-mode emission threshold at fixed $\tau$.
        }
		\label{fig:scale}
	\end{center}
\end{figure*}

The origin of fast radio bursts (FRB) is yet an open issue.   The seemingly distinct behavior of repeating FRBs and their non-repeating
counterparts, suggests that it is quite likely that more than one class of astrophysical objects generate these bursts. 
It has been widely thought that the class of repeaters is associated with magnetars - strongly magnetized neutron stars. 
The recent detection of a FRB from a known Galactic magnetar (although at the faint end of the population \cite{Bochenek2020}) provides support 
to the hypothesis that magnetars can produce FRB \citep[for review see][]{lyubarsky_emission_2021}. 

Relativistic magnetized shocks have emerged as plausible emission sites of coherent radio bursts that may account for the observed FRB \citep{lyubarsky_model_2014,waxman_origin_2017,metzger_fast_2019,iwamoto_precursor_2019,plotnikov_synchrotron_2019,vanthieghem_fast_2025}.
Such shocks can form in the inner magnetosphere of a magnetar upon abrupt steepening of a millisecond fast magnetosonic wave \citep{chen_propagation_2022,beloborodov_monster_2023,vanthieghem_fast_2025,Bernardi2025,Grehan2026}, or in shocks
that form by the interaction of an escaping magnetosonic pulse with the surrounding medium well beyond the light cylinder \citep{lyubarsky_model_2014,waxman_origin_2017,metzger_fast_2019}.
Recent plasma kinetic simulations have revealed inherent generation of a high frequency precursor wave 
within a nonlinear solitary structure that form at the foot of the shock transition layer \citep{iwamoto_precursor_2019,plotnikov_synchrotron_2019,sironi_coherent_2021,vanthieghem_fast_2025,Bernardi2025}.
The details of the underlying emission mechanism of the precursor wave have not been fully explored yet, although 
it likely involves the synchrotron maser instability, owing to formation of momentum ring distributions in the plasma 
crossing the solitons, prior to the onset of dissipation. 
However, the presence of harmonics associated with the distance between consecutive solitons in the spectrum of the precursor wave suggests that additional factors are at play.

In a previous paper \citep{vanthieghem_relativistically_2025} we constructed an analytic model of magnetized shocks, that treats the plasma as  a large collection of stationary 
cold beams coupled through the magnetic field, and used it to compute the  kinetic-scale shock transition.
We have shown that the  model can robustly reproduce the global shock structure observed in the Particle-In-Cell (PIC) simulations and, in addition,
elucidates the underlying physics.   In particular, we demonstrated  that shock dissipation does not require collective plasma interactions, as in the 
case of unmagnetized shocks, rather it stems
from nonlinear dynamics of charged particles in a stationary, quasiperiodic system of collisionless fast-magnetosonic
solitons.   The detailed shock structure is controlled by two parameters: the upstream Mach number and temperature. 
However, our analysis ignored the effect of radiation reaction on the shock structure, which might be important in monster shocks \citep{beloborodov_monster_2023}, and conceivably also in far-wind models under certain conditions. 

The primary goal of this paper is to study the effect of strong cooling on the generation of precursor-wave emission in strongly magnetized shocks, particularly monster shocks. To this end, we employ the \entity~PIC code~\citep{Hakobyan2025,Boss_2026}, adapted to the strong-cooling regime, in combination with the multifluid shock model described in \cite{vanthieghem_relativistically_2025}. Here, we generalize the multifluid model by incorporating the radiative backreaction force into the fluid equations. Combining PIC simulations with semi-analytic solutions of the shock equations enables us to derive approximate scaling relations for the shock properties that can be extrapolated to the regime expected in astrophysical systems. We find that monster shocks remain highly supercritical even under strong cooling. Consequently, cooling does not quench the precursor-wave emission, regardless of the luminosity of the driving kHz wave. The luminosity of the FRB emitted by a monster shock is therefore determined by the luminosity of the kHz wave and the precursor-wave generation efficiency, $\epsilon\sim10^{-3}$.

The paper is organized as follows.
In Sec.~\ref{sec:theory}, we first examine the shock structure and its associated scaling laws using the generalized $N$-fluid model introduced in~\cite{vanthieghem_relativistically_2025}. In Sec.~\ref{sec:PIC}, we investigate the impact of radiative cooling on the resulting coherent emission as a function of the cooling strength parameter, combining the fluid framework with fully kinetic numerical simulations. Finally, in Sec.~\ref{sec:monster}, we apply these results to the astrophysically relevant case of monster shocks propagating in magnetar magnetospheres as fast radio burst progenitors.

\section{Stationary Shock Structure \label{sec:theory}}

In this section, we compute the general structure of relativistically magnetized shock waves propagating in a symmetric pair plasma, in the regime of strong cooling.
We do so by employing a modified multifluid shock model that treats the plasma as a collection of $2N$ cold beams, with 
one set of $N$ beams accounts for electrons and one for positrons.
The assumption that the
pair plasma is symmetric ensures that it is sufficient to solve only the equations of motion of the  positron beams; the solution
for the electron beams readily follows from the symmetry \citep{vanthieghem_relativistically_2025}.  
However, the contribution of both 
species must be properly included in the source terms of the electric and magnetic fields, $\bm{E}$ and $\bm{B}$.  
We stress that the symmetry
of the electrons and positrons' trajectories is preserved also in the presence of a radiation reaction force, as can be inferred
from Eq.~\eqref{eq:RRforce} below \footnote{One can verify that $\mathcal{F}^i_x \to \mathcal{F}^i_x$ and $\mathcal{F}^i_y \to -\mathcal{F}^i_y$ under
the transformation $(u^i_x, u^i_y) \to (u^i_x, -u^i_y)$, as required by the symmetry of $e^\pm$ trajectories.}. 
The $i$th beam of positrons is characterized by a proper
density $n^i$ and 4-velocity $\bm{u}^i = \gamma^i \bm{\beta}^i$, where the index $i = 1,\dots,N$ label the beam number. 
The equation of motion of the $i$th positron beam reads:
\begin{equation}
(\partial_t + \bm{\beta}^i\cdot\nabla)\bm{u}^i= e(\bm{E} + \bm{\beta}^i\times \bm{B}) +  \bm{\mathcal{F}}^i,
\end{equation}
where
\begin{equation}
    \bm{\mathcal{F}^i}\,=\,\frac{\sigma_{\rm T} B_0^2}{4 \pi\,\gamma_0^2} \, \left[ \bm{\mathcal{\kappa}^i} - (\gamma^i)^2 (\chi_{\rm R}^i)^2 \bm{\beta}^i \right] 
       \label{eq:RRforce}
\end{equation}
is the radiation reaction force acting on the beam, $e>0$ is the positron charge,  $\sigma_{\rm T}$ is the Thomson cross section, 
and $\bm{\mathcal{\kappa}^i}$ and $\chi_{\rm R}^i$ are dimensionless quantities defined as:
\begin{align}
    &\bm{\kappa^i}\,=\, \left( \bm{\tilde{E}} + \bm{\beta}^i \times \bm{\tilde{B}} \right) \times \bm{\tilde{B}}\,+\,\left(\bm{\beta}^i \cdot \bm{\tilde{E}}\right)\,\bm{\tilde{E}}\,, \label{eq:kappa}\\
    &(\chi_{\rm R}^{i})^2\,=\, \left( \bm{\tilde{E}} + \bm{\beta}^i \times \bm{\tilde{B}} \right)^2 - \left(\bm{\beta}^i \cdot \bm{\tilde{E}}\right)^2 \,,\label{eq:chi2}
\end{align}
with $\bm{\tilde{B}} = \gamma_0 \bm{B} / B_0$ and $\bm{\tilde{E}} = \gamma_0 \bm{E} / B_0$ representing the electromagnetic fields normalized to a proper fiducial background magnetic field $B_0 / \gamma_0$.  

Following~\citep{vanthieghem_relativistically_2025}, we assume that the shock structure is one-dimensional and stationary, with the upstream flow moving in 
the $+\hat{x}$ direction and the magnetic field perpendicular to the flow along the $z$ direction.  
The Maxwell–Faraday equation then implies that the electric field is constant, $\bm{E} = B_0\beta_0 \hat{y}$ ($\tilde{E} = u_0$).  The 
electric current density is given by $\bm{j} = j_y \,\hat{y}$, with $j_y= 2e \sum_i n^i u^i_y$. 
The equations are rendered dimensionless by normalizing velocities to the upstream velocity $u_0$, $\tilde{u}^i_x = u^i_x/u_0$, $\tilde{u}_y^i = u_y^i/u_0$,
and measuring length in units of the skin depth, $\tilde{x} = x \, \omega_{\rm p}/c$, where $\omega_{\rm p} = \sqrt{4\pi e^2 n_{\rm up}/m_e }$ is the upstream 
plasma frequency, $n_{\rm up}\,=\,2 n_0$, the total plasma density, and $n_0 =\sum_i n^i(0)$ is the total proper positron density of the upstream flow.
With the above considerations, the set of equations governing the shock transition reads:
\begin{align}
    &\partial_{\tilde{x}} \tilde{u}^i_x \,=\, \frac{\tilde{B}_z}{\Ma}\,\frac{\tilde{u}^i_y}{\tilde{u}^i_x} + \frac{\tau \, u_0^4}{\Ma^4} \left[ \tilde{u}_y^{i\,2} + \right. \nonumber\\ 
    &\qquad\left. \frac{\tilde{B}_z \gamma^i}{u_0^2}\,(2 \tilde{u}^i_x+ \frac{1}{u_0^2 \tilde{u}^i_x}) - \left( \frac{\tilde{B}_z^2}{u_0^2} + 1 \right) \frac{\gamma^{i\,2}}{u_0^2} \right] \,,\label{eq:1_fluid_1}\\ 
    &\partial_{\tilde{x}} \tilde{u}^i_y \,=\, - \frac{\tilde{B}_z}{\Ma} +  \frac{1}{\Ma} \frac{\gamma^i}{\tilde{u}^i_x} \nonumber \\ 
    &\qquad\qquad- \frac{ \tau \, u_0^4}{\Ma^4}\frac{\tilde{u}^i_y}{\tilde{u}^i_x}\left( \tilde{u}^i_x - \frac{\tilde{B}_z \gamma^i}{u_0^2} \right)^2\,,\label{eq:1_fluid_2}\\ 
    &\partial_{\tilde{x}} \tilde{B_z} \,=\, - \Ma\,\frac{1}{N}\,\sum^N_i\,\frac{\tilde{u}^i_y}{\tilde{u}^i_x}\,, \label{eq:1_fluid_3} 
\end{align}
where $M_A= u_0/\sqrt{\sigma}$ is the Alfvénic Mach number, $\sigma$ is the magnetization of the upstream flow, and
\begin{align} \label{eq:tau}
\tau = \frac{\sigma_{\rm T}\,n_{\rm up}\,c}{\omega_{\rm p}} \, u_0 \Ma^2
\end{align}
is a dimensionless cooling parameter\footnote{The cooling parameter discussed here is exactly equivalent to the parameter $\chi$ introduced in Eq. (B29) of~\citep{beloborodov_monster_2023}, where the general jump conditions are derived. We adopt the notation $\tau$ to avoid confusion with the quantum parameter.}, with its normalization discussed below.
In a strongly magnetized relativistic system, $\bm{\kappa}$ is often neglected. However, in the case considered here, we must include the full expression to obtain the correct downstream compression factor.  

Figure~\ref{fig:scale} shows the solutions of Eqs.~\eqref{eq:1_fluid_1}–\eqref{eq:1_fluid_3} for different values of the cooling parameter $\tau$. For weak cooling, $\tau \ll 1$, the beam motion remains coherent during the first few soliton crossings, before complete phase mixing sets in. This phase mixing dissipates the energy and erases the solitary structure, as previously found for uncooled plasmas \citep{vanthieghem_relativistically_2025}. As the cooling strength increases, the amplitude of the solitary wave decreases, while its coherence increases. Since the emission of the precursor wave depends on both the amplitude and the coherence of the gyrational beam motion, its dependence on the cooling strength is expected to be non-linear, as the competing effects of amplitude reduction and increased coherence partially counteract each other. 

As also shown in Fig.~\ref{fig:scale}, the wavelength of the solitary wave decreases with increasing cooling strength, implying that the frequency of the precursor wave should increase accordingly. Ultimately, in the regime of excessive cooling, \emph{i.e.}, when the cooling length becomes comparable to or shorter than the gyroradius, the solitary structure, and hence the precursor wave emission, are completely quenched. These trends are confirmed by direct PIC simulations presented in Section~\ref{sec:PIC}.

To quantify the trends identified above, we now focus on the scaling of the cooling timescale. Equation~\eqref{eq:tau} defines a fiducial dissipation parameter motivated by an analytical estimate obtained in the strongly relativistic regime where $\tilde{u}_y \sim 0$, $\gamma \simeq \tilde{u}_x u_0 + \frac{1}{2 \tilde{u}_x u_0} \gg 1$, and considering the dynamics of a single fluid, that is, neglecting stochastic dissipative effects. In this limit, and to leading order, energy conservation in the presence of radiative cooling takes the form:
\begin{align}
    \partial_x \left( \frac{u_0 \delta\tilde{B}_z}{\Ma^2}  + \frac{\gamma}{u_0} \right) \,\simeq\, -\,\tau\,\left( \frac{u_0 \delta\tilde{B}_z}{\Ma^2} \right)^2\, \left(\frac{\gamma}{u_0}\right)^2,
\end{align}
assuming $\delta\tilde{B}_z\,\gg\,1/u_0$, where $\delta\tilde{B}_z\,\equiv\,\tilde{B}_z - \gamma_0$. The evolution equation for $\gamma$, needed to close the set, also involves the combination $u_0 \delta\tilde{B}_z/\Ma^2$ in the leading order.

From this we directly obtain the rescaling of the magnetic field compression: $\delta \tilde{B}_z^{\rm max}(\tau) \sim \alpha(\tau)\, \Ma^2/u_0$, or $\delta B_z^{\rm max}(\tau)/B_0 \sim \alpha(\tau) \Ma^2/u_0^2 = \alpha(\tau) /\sigma$, where $\alpha(\tau)$ is of order unity for reasonable values of $\tau$. The present analytical scaling is therefore valid in the regime $1 \ll \Ma \ll \sqrt{u_0}$. 

We demonstrate the validity of this scaling in the strongly nonlinear regime through direct numerical integration. Figure~\ref{fig:scale} presents the shock profiles obtained for increasing cooling strengths, $\tau\,=\,[10^{-3},\,10^{3}]$, for several combinations of shock velocity, $u_0\,=\,10^3,\,10^4$, and Alfv\'enic Mach number, $\Ma\,=\,30,\,100$. As illustrated, both the peak solitary compression at the shock transition and the characteristic cooling envelope of the downstream soliton train exhibit a consistent scaling across the explored parameter space. The remaining dependence on the shock parameters is primarily controlled by the logarithmic $\Ma$-dependence of the soliton period, as previously identified in \citet{vanthieghem_relativistically_2025}. In the next section we discuss the exact scaling of the compression.

In short, the structure of a relativistically magnetized pair shock in the strong-cooling regime is captured by a stationary multifluid model in which the plasma is represented by $N$ cold beams evolving under the full radiation reaction force of Eq.~\eqref{eq:RRforce}. The transition is then controlled by two dimensionless numbers only: the Alfv\'enic Mach number $\Ma$ and the cooling parameter $\tau$ of Eq.~\eqref{eq:tau}. Increasing $\tau$
lowers the amplitude of the solitary structure while increasing its coherence and shortening its wavelength, and, once the cooling length falls below the gyroradius, the solitary structure is quenched altogether.

\section{Coherent emission}
\label{sec:PIC}

\begin{table}[t]
\centering
\renewcommand{\arraystretch}{1.3}
\begin{ruledtabular}
\begin{tabular}{lccccc}
 & $\mathcal{S}_1$ & $\mathcal{S}_2$ & $\mathcal{S}_3$ & $\mathcal{S}_4$ & $\mathcal{S}_5$ \\
\hline
$\kappa_0$               & $10^{-7}$          & $10^{-5}$ & $2.5\times10^{-4}$ & $10^{-3}$ & $6.25\times10^{-3}$ \\
$\tau_{\rm max}$         & $6.8\times10^{-3}$ & $0.34$    & $4.4$              & $11$      & $2.5\times10^{3}$  \\
$\Ma/\Ma^{\rm crit}$     & $10$               & $3.0$     & $0.8$              & $0.5$     & $0.2$              \\
\end{tabular}
\end{ruledtabular}
\caption{Simulation parameters: Each simulation is characterized by a fixed fiducial background cooling factor
$\kappa_0\,=\,\sigma_{\rm T} n_{\rm bg} c / \omega_{\rm p,bg}$. Owing to the shock evolution, the effective cooling parameter $\tau$ varies in time. We list $\tau_{\rm max}$, the maximum value of $\tau$ attained in each simulation, and the corresponding ratio between the Alfv\'enic Mach number and the critical Alfv\'enic Mach number showing the transition to the subcritical regime. The case $\mathcal{S}_0$, for which radiative losses are neglected ($\kappa\,=\,\tau_{\rm max}\,=\,0$), is omitted from the table. \label{tab:sims}}
\end{table}.

\begin{figure}[t]
	\begin{center}
		\includegraphics[width=1\columnwidth]{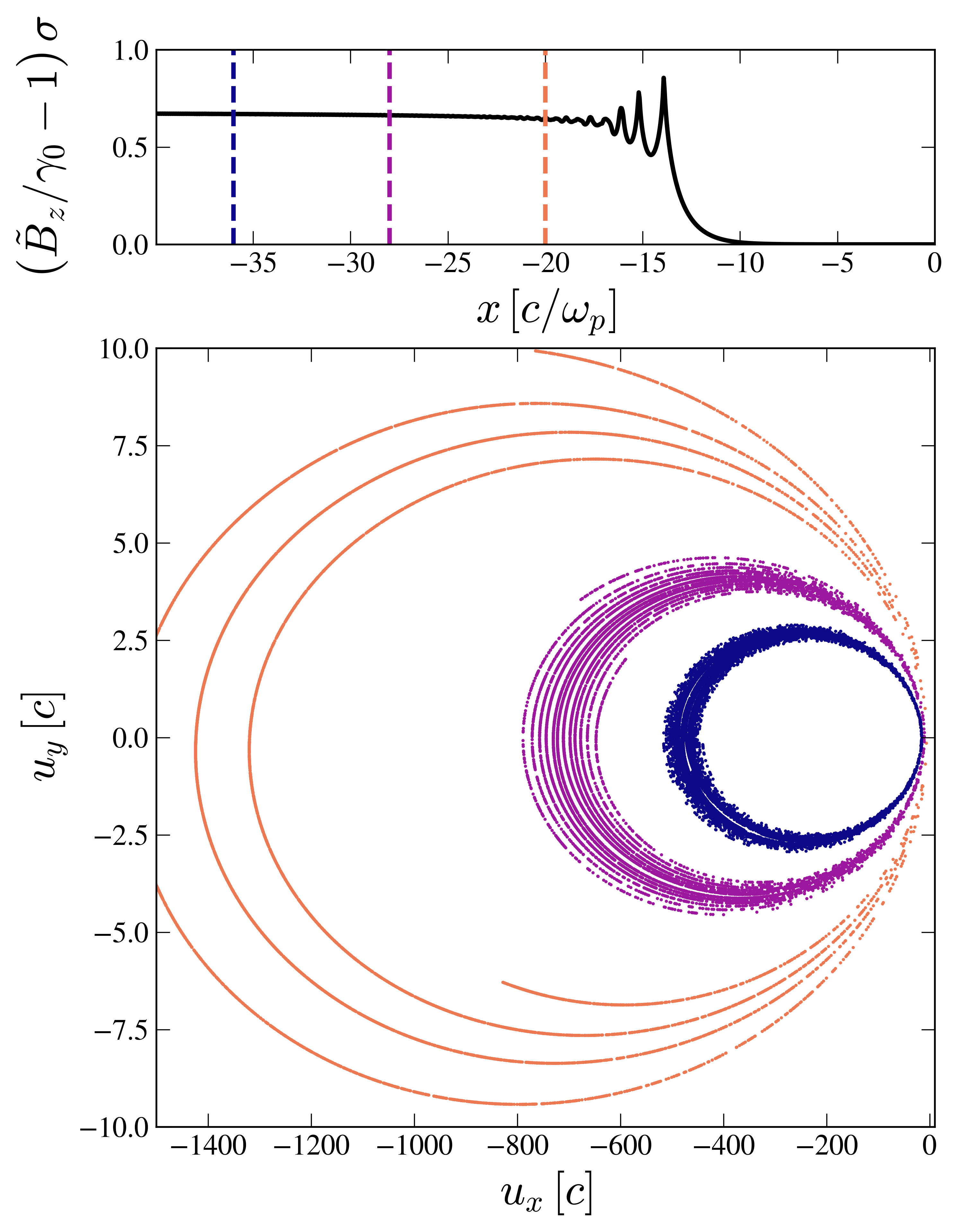}
		\caption{The top panel shows the normalized magnetic-field compression ratio profile for $N = 10^4$, $\tau\,=\,1$, $u_0 = 10^4$, and $M_{\rm A} = 100$. The bottom panel illustrates the corresponding phase-space velocity distribution of the $10^4$ fluid beams evaluated at the positions indicated by the vertical dashed lines in the top panel.
        }
		\label{fig:phase_space}
	\end{center}
\end{figure}

In the previous section, we saw that radiation reaction modifies the shock structure, with $\tau$ as the controlling parameter.  In \citet{vanthieghem_relativistically_2025} we showed that the leading-order
structure of relativistically magnetized shocks is not set by the unstable modes responsible for the emission of extraordinary ($X$) modes through the synchrotron-maser instability.  That conclusion rests on weak coupling between the upstream plasma and the forward-emitted wave, an assumption that is not always satisfied here, as discussed in Sec.~\ref{sec:monster}.  \citet{Zhang_2025} report that cooling alters the emissivity of such a shock and can enhance it, albeit over a narrow range of parameters.  We are still missing a general scaling of the emission threshold in terms of the two shock parameters, $\tau$ and $\Ma$, which we derive below.

The synchrotron-maser instability is driven by ring-type distributions, and its growth rate depends strongly on both the radius of the ring in the $(\tilde{u}_\parallel,\tilde{u}_\perp) = (\tilde{u}_x,\tilde{u}_y)$ plane and on its annular width, \emph{i.e.} the ring temperature.  Cooling acts on these two quantities in opposite ways: it reduces the ring radius, lowering the growth rate, and it narrows the annulus, raising it.  The net dependence on $\tau$ is therefore non-monotonic, and the emission is expected to be most efficient at intermediate cooling strength.  Figure~\ref{fig:phase_space} illustrates the effect on the phase-space distribution for $\tau\,=\,1$, $\Ma\,=\,100$ and $u_0\,=\,10^4$, using $N\,=\,10^4$ fluid beams.  In contrast to the uncooled case of \citet{vanthieghem_relativistically_2025}, in which phase mixing rapidly broadens the ring, the distribution here remains cold over extended downstream distances.  We stress that these are genuine rings in the local velocity distribution at fixed $\tilde{x}$, and not the apparent rings obtained by averaging a helical trajectory over several gyration periods.  Such a configuration is favorable to the development of population inversion and to the associated coherent emission \citep{hoshino_preferential_1991, Bilbao_2025}.

\begin{figure}[t]
	\begin{center}
		\includegraphics[width=1\columnwidth]{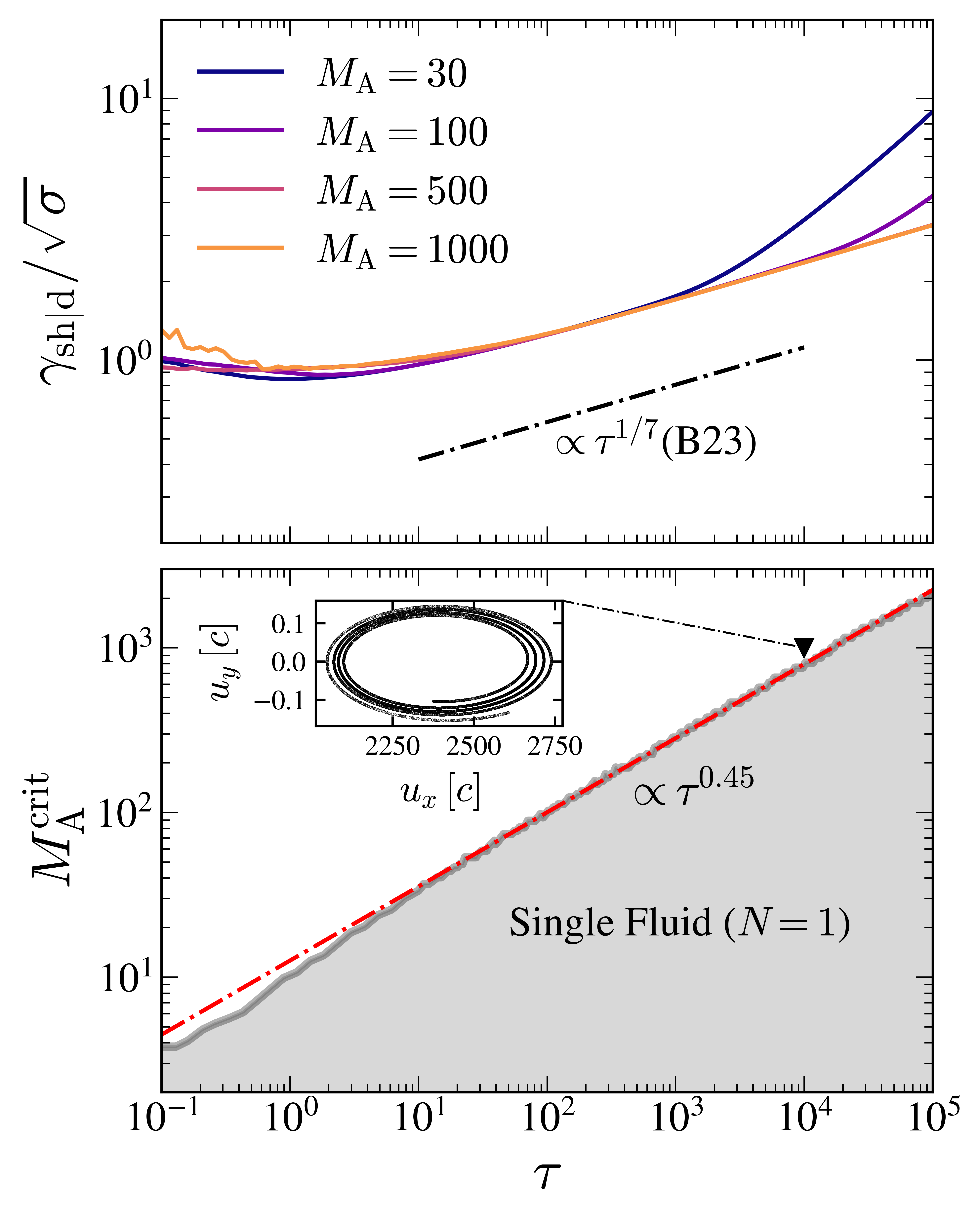}
        \caption{
        \emph{Top:} Shock compression versus the cooling parameter $\tau$ for
        $\Ma=30$, $100$, $500$, and $1000$. Over $3\lesssim\tau\lesssim10^3$,
        all four curves collapse onto the scaling $\propto\tau^{1/7}$ (black
        dash-dotted line), corresponding to the radiative jump condition of
        \citet{beloborodov_monster_2023} (B23). Once $\Ma$ falls below the
        $\tau$-dependent critical value $\Ma^{\rm crit}(\tau)$, the shock becomes
        subcritical and the compression departs from this scaling.
        \emph{Bottom:} Critical Alfv\'enic Mach number $\Ma^{\rm crit}$ for the
        onset of chaotic beam dynamics as a function of $\tau$, obtained by
        comparing multifluid solutions with $N=5$ with the single-fluid solution over
        a $10^2\times10^2$ grid in $(\tau,\Ma)$. In the shaded region, the
        multifluid solutions are laminar and coincide with the single-fluid
        solution ($N=1$): phase mixing is negligible, no ring distribution
        forms, and no significant coherent emission is expected. The red
        dash-dotted line shows the fit
        $\Ma^{\rm crit}\simeq12.6\,\tau^{0.45}$.
        \emph{Inset:} Immediate downstream phase-space distribution in the
        $(u_x,u_y)$ plane for the point marked by the triangle on the threshold
        curve ($\tau\simeq10^4$, $\Ma\simeq10^3$, and $u_0=10^6$), computed
        using $N=10^4$.
        }
    	\label{fig:MAcrit}
	\end{center}
\end{figure}

Downstream of the transition, the ring relaxes at a rate set by the
jump conditions, which are modified by cooling.  \citet{beloborodov_monster_2023} showed that the shock Lorentz factor relative to the downstream plasma scales as
$\gamma_{\rm sh|d} \simeq (1+\tau)^{1/7}\sqrt{\sigma_{\rm u}}$.  The multifluid model recovers this scaling, where solutions with different Mach numbers collapse onto a single curve (top panel of Fig.~\ref{fig:MAcrit}).  At larger $\tau$ they depart from it earlier and reach larger values for smaller compression at lower $\Ma$. Recovering the correct jump conditions before and near this transition requires retaining the $\bm{\kappa}$ term of Eq.~\eqref{eq:RRforce}, which is usually neglected in the strongly magnetized limit. As can be seen, this regime corresponds to a full suppression of the solitary structure, for which no emission is expected. However, efficient emission of the precursor wave is suppressed already when the flow approaches the subcritical regime, before complete quenching of the solitary structure. 

If the emission of $X$ modes requires the formation of a localized ring distribution, then a necessary, but not sufficient, condition is the onset of chaotic beam trajectories associated with the supercritical regime of the shock, and the ensuing phase mixing.  Without radiation reaction, chaos sets in only above a critical Mach number, $\Ma > \Ma^{\rm crit} \simeq 2$ \citep{vanthieghem_relativistically_2025}; below it the shock transition is subcritical and the multifluid solutions collapse onto the single-fluid solution.  We determine the dependence of $\Ma^{\rm crit}$ on $\tau$ by scanning $10^2$ values of $\tau \in [10^{-1},10^5]$ and $10^2$ values of $\Ma \in [2,10^4]$ with $N\,=\,5$ beams, and measuring the integrated departure of the multifluid field profile from the single-fluid one,
$\mathcal{D} = \sigma^2\int (\tilde{B}_z^{N=1} - \tilde{B}_z^{N = 5})^2\,{\rm d}\tilde{x}/\gamma_0^2$.
A sharp transition occurs at $\Ma^{\rm crit}$: $\mathcal{D}$ either takes an $O(1)$ value or falls to the numerical floor, for which the solutions are degenerate.  As discussed in~\citet{vanthieghem_relativistically_2025}, the critical transition is insensitive to $N$, as long as $N\geq2$.  As illustrated in the bottom panel of Fig.~\ref{fig:MAcrit}, a fit over the range of non-negligible cooling gives $\Ma^{\rm crit} \simeq 12.6\,\tau^{0.45}$. A continuous transition between the regimes of supercritical and subcritical shocks therefore scales as:
\begin{align}
\Ma^{\rm crit} \simeq 2\,+\, 12.6\, \tau^{0.45}.    
\label{eq:M_A_crit}
\end{align}
Significant coherent $X$-mode emission is expected only above this critical point.

\begin{figure}[t]
	\begin{center}
		\includegraphics[width=1\columnwidth]{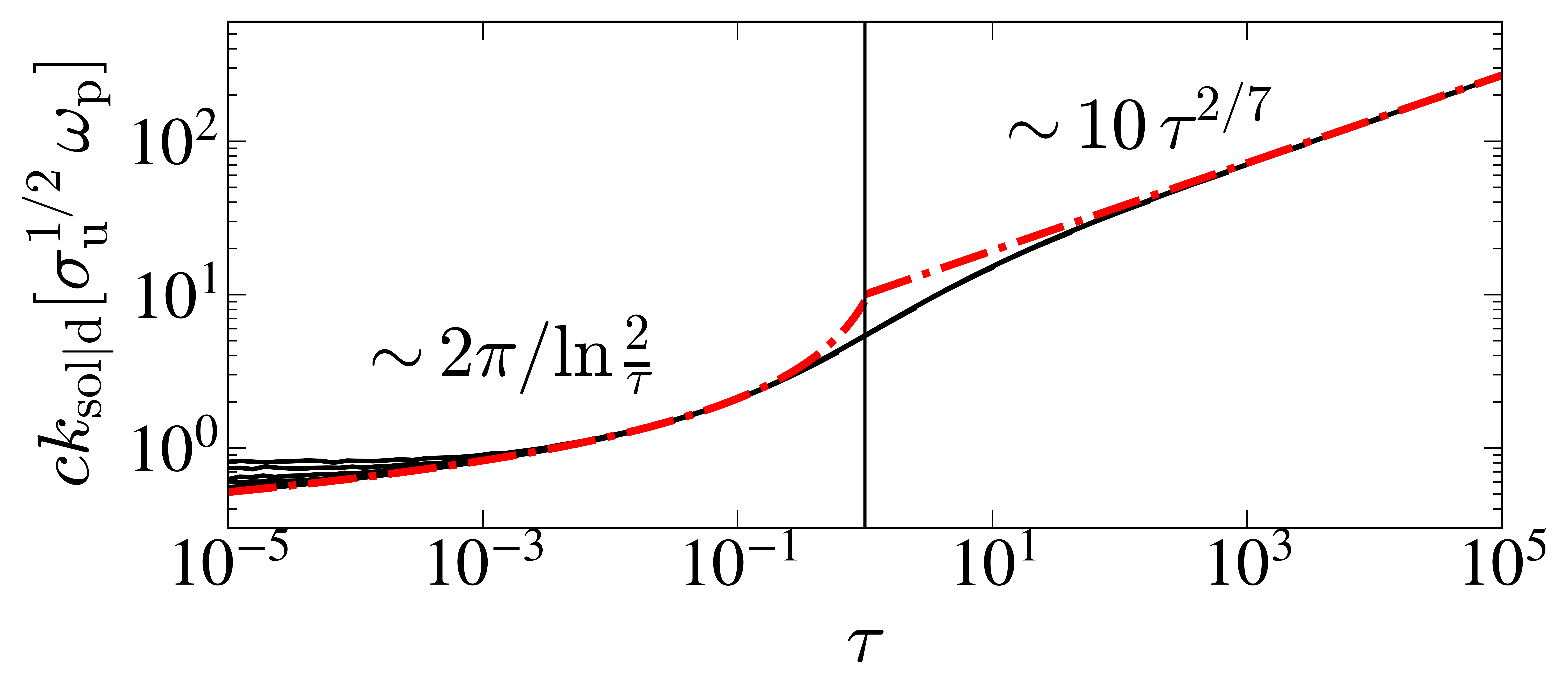}
		\caption{
        Downstream-frame soliton wavenumber, $k_{\rm sol|d}=\gamma_{\rm sh|d}k_{\rm sol|sh}$, as a function of the
        cooling parameter $\tau$. Black curves show the multifluid solutions for $\Ma\in[10,\,3\times10^3]$ at $u_0=10^7$, each truncated at
        $\Ma^{\rm crit}(\tau)$. The red dot-dashed curve shows the analytical estimate, with the vertical line marking the transition between its two
        branches.}
		\label{fig:ksol}
	\end{center}
\end{figure}

Above this threshold, emission is expected at a peak frequency set by the typical spacing between successive solitons, which acts as a resonant cavity~\citep{plotnikov_synchrotron_2019}.  An estimate of the cavity size in the shock front frame, $\lambda_{\rm sol|sh}$,  for an uncooled shock was obtained in~\citet{vanthieghem_relativistically_2025} and depends only logarithmically on $\Ma$.  That estimate extends naturally to a cooled shock. For $\Ma \gg 1$, we obtain for the soliton wavenumber in the shock-front frame, $k_{\rm sol|sh}\,=\,2 \pi/\lambda_{\rm sol|sh}$:
\begin{align}
\frac{c k_{\rm sol|sh}}{\omega_{\rm p}} \simeq
\begin{cases}
\; 2\pi \left/ \ln\!\left( \dfrac{4}{2\tau + \Ma^{-4}} \right) \right. & \quad \left(\tau < 1\right)\,,\\[6pt]
\; 10\,\tau^{1/7} & \quad \left(\tau > 1\right)\,.
\end{cases}
\end{align}
The relevant wavenumber is the one measured in the downstream frame.  Treating the soliton train as a static structure swept past the downstream plasma in the shock front frame, $k_{\rm sol|d} = \gamma_{\rm sh|d}\,k_{\rm sol|sh}$, which we take as a proxy for the $X$-mode peak frequency in that frame.  A comparison between the numerical integration and this estimate is shown in Fig.~\ref{fig:ksol} where good agreement is found both in the regime of weak and strong cooling. For $\tau \gg 1$, we therefore expect:
\begin{align} \label{eq:peakom}
    \omega_{|d} \sim c k_{\rm sol|d} \simeq 10\,\tau^{2/7}\sqrt{\sigma_{\rm u}}\,\omega_{\rm p}\,.
\end{align}
Stronger cooling thus pushes the peak emission progressively deeper into the kinetic regime of the wave.

As a stationary description, the multifluid model does not capture $X$-modes, associated with plasma instabilities. To quantify the coherent emission from the shock transition, we perform a suite of Particle-In-Cell (PIC) simulations using the massively parallel, hardware-agnostic code \entity~ \citep{Hakobyan2025,Boss_2026}. This section focuses on the emission efficiency, while the global dynamics of the monster shock are addressed in the following section. For numerical efficiency, we follow wave steepening in the background-plasma frame within a window moving at the speed of light~\citep{Vay_2020}.

A cold, magnetized plasma is injected at rest with constant density through the right boundary of a moving window propagating at the speed of light, temperature $k_{\rm B}T\,=\,10^{-4}\,\me c^2$, and a prescribed magnetization profile $\sigma\,=\,\sigma_0\left(1+x/\mathcal{R}_0\right)^{-3}$. The wavelength is set to $\lambda_{\rm w}\,=\,1000\,c/\wpp\,=\,0.1\,\mathcal{R}_0$. The wave is launched with amplitude $0.3\,B_0$ in a plasma with $\sigma_0\,=\,2.8\times 10^4$, ensuring steepening at $x_{\rm s}\,\simeq\,4060\,c/\wpp$, where the magnetization is $\sigma_{\rm s}\,=\,10^4$. We integrate radiation reaction via Strang splitting with a fourth-order Runge–Kutta scheme. For each simulation, we use a resolution $\Delta x \,=\, 0.0024\,c/\omega_{\rm p, bg}$, with $\omega_{\rm p, bg}$ the plasma frequency of the injected background plasma, $c\Delta t \,=\, 0.5 \Delta x$, and 200 particles per cell.

To investigate the scaling and feedback associated with radiative losses, we impose a fiducial background cooling factor, $\kappa_0\,=\,\frac{\sigma_{\rm T} n_{\rm bg} c}{\omega_{\rm p,bg}}\,=\,\tau |u_{\rm u|sh}|^{-1}\Ma^{-2}\left(n_{\rm bg}/n_{\rm up}\right)^{1/2}$, where $n_{\rm bg}$ denotes the total pair constant background density upstream of the wave, $n_{\rm up}$ is the proper density immediately upstream of the shock, and $u_{\rm u|sh}\,\equiv\,u_0$ is the upstream four-velocity measured in the shock frame (Sec.~\ref{sec:theory}). We present six simulations  with $\kappa_0\,=\,[0,\,10^{-7},\,10^{-5},\,2.5\times10^{-4},\,10^{-3},\,6.25\times10^{-3}]$ held fixed in time. As a result of the intrinsic monster-shock evolution, the effective cooling parameter $\tau$ varies throughout the steepening (Sec.~\ref{sec:monster}). The values of the cooling parameter $\kappa_0$ are listed in Tab.~\ref{tab:sims}, together with the corresponding maximum value of the scaled cooling factor $\tau_{\rm max}$.

Figure~\ref{fig:EM_sh} illustrates the shock dynamics as a function of the cooling parameter $\tau$. One of the most striking effects is the upshift of the peak emission frequency. For $\mathcal{S}_4$ ($\sigma_{\rm u}\simeq180$, $\tau\simeq4.0$), the multifluid integration shown in Fig.~\ref{fig:ksol} gives $\omega_{\rm |d}\simeq107\,\omega_{\rm p}$, consistent with the peak of the bottom-right panel of Fig.~\ref{fig:EM_sh}. The asymptotic form of Eq.~\eqref{eq:peakom} yields $\omega_{\rm |d}\,\simeq200\,\omega_{\rm p}$.
Although $\Ma$ and $\Ma^{\rm crit}$ both evolve during the steepening, their ratio stays nearly constant within each run, taking the values $\Ma/\Ma^{\rm crit} \simeq [18,\,10,\,3.0,\,0.8,\,0.4,\,0.2]$ for $\mathcal{S}_0$--$\mathcal{S}_5$ (Tab.~\ref{tab:sims}), measured in the immediate shock upstream. The threshold is thus crossed between $\mathcal{S}_2$ and $\mathcal{S}_3$: the three weakly cooled runs are supercritical, while the three strongly cooled runs are in or near the subcritical regime. The progressive suppression of $X$-mode emission with increasing $\tau$ therefore correlates with the fall of $\Ma/\Ma^{\rm crit}$ below unity. Near threshold, $\Ma\simeq\Ma^{\rm crit}$, the emission becomes strongly narrow-banded. Because $\Ma^{\rm crit}\,\to\,2$ as $\tau\,\to\,0$, the uncooled shock at $\Ma\simeq2$ realizes the same condition, and the same spectral properties are recovered there. Narrow-banding is therefore expected to signal the transition to the subcritical regime on top of the cooling.

\begin{figure}[t]
	\begin{center}
		\includegraphics[width=1\columnwidth]{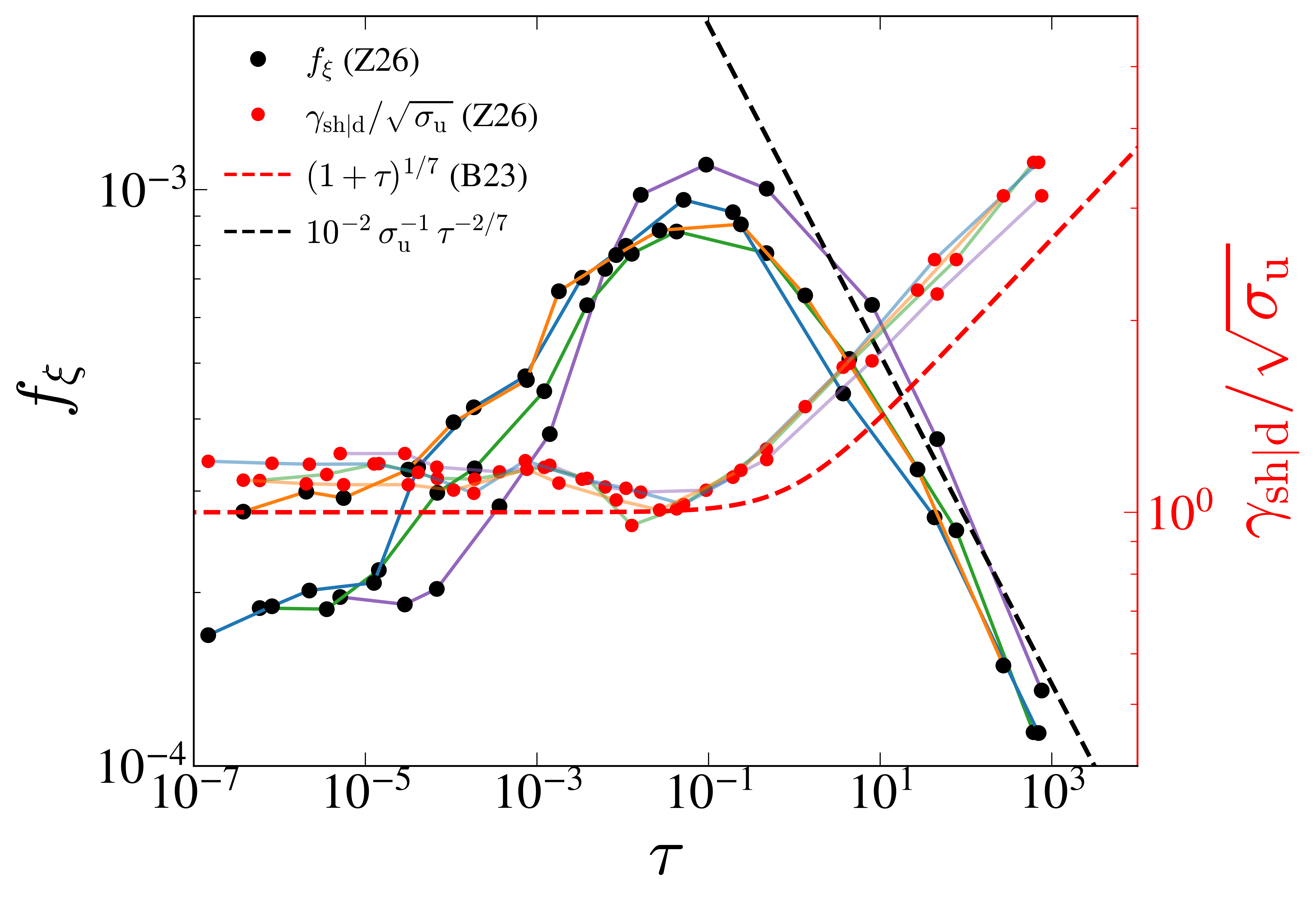}
        \caption{
        Emission efficiency and shock Lorentz factor from the simulations of~\citet{Zhang_2025} (Z26), recast in terms of the cooling parameter $\tau$ (simulation parameters in their Fig.~S3). \emph{Black, left axis:} emission efficiency $f_\xi$; the black dashed line shows $f_\xi = 10^{-2}\,\sigma_{\rm u}^{-1}\,\tau^{-2/7}$. \emph{Red, right axis:} shock Lorentz factor in the downstream frame, normalized to the upstream magnetization, $\gamma_{\rm sh|d}/\sqrt{\sigma_{\rm u}}$; the red dashed line shows the radiative jump condition $(1+\tau)^{1/7}$ of~\citet{beloborodov_monster_2023} (B23). Coloured lines connect runs with the same background density; all runs have $\sigma_{\rm u} = 10$.}
		\label{fig:zhang}
	\end{center}
\end{figure}

\begin{figure*}[t]
	\begin{center}
		\includegraphics[width=1\textwidth]{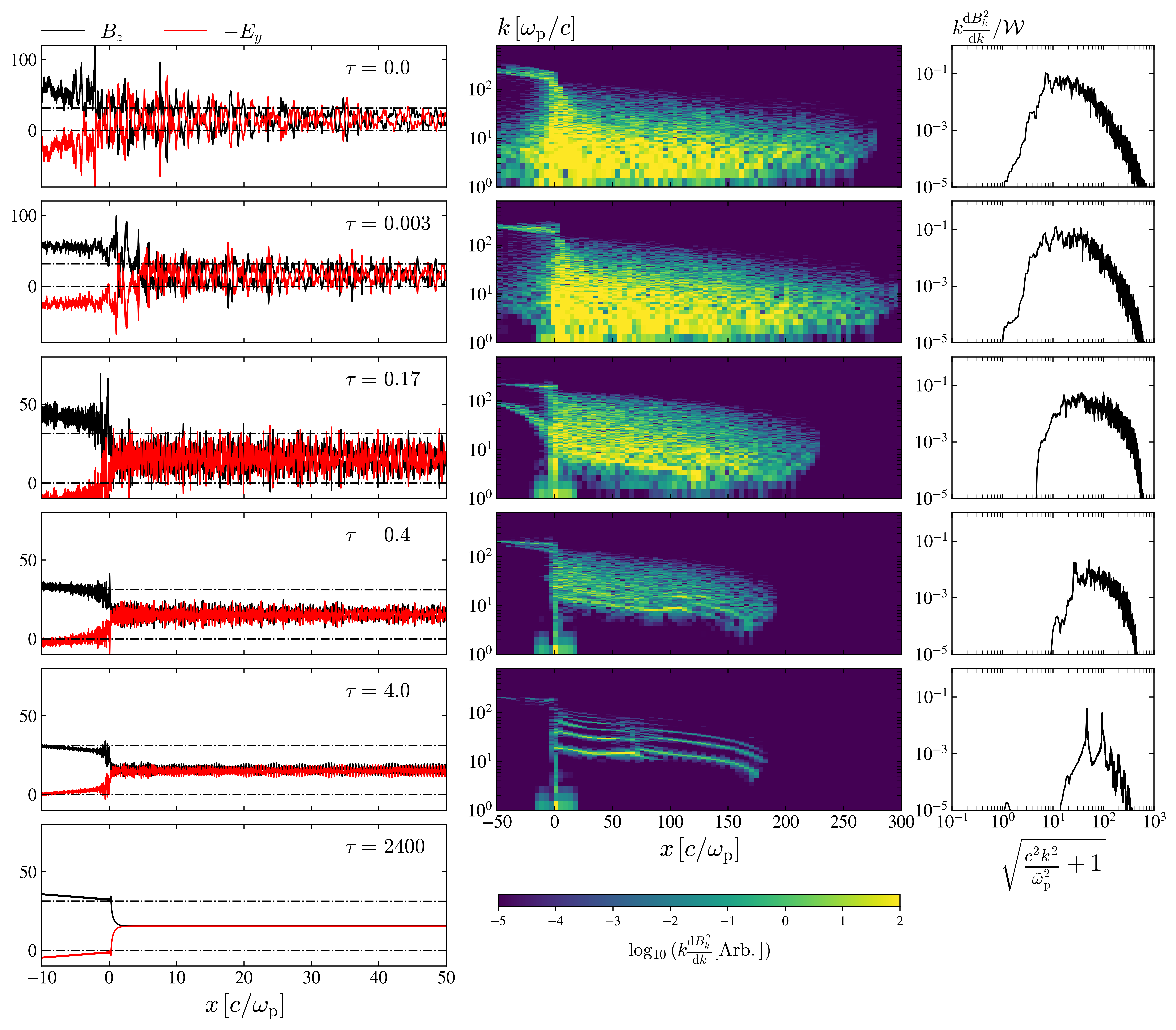}
		\caption{Profiles of the electromagnetic fields $B_z$ (black) and $-E_y$ (red) at $\wpp t = 9750$ for $\frac{\sigma_{\rm T} n_{\rm bg} c}{\omega_{\rm p,bg}} = [10^{-5},\, 2.5 \times 10^{-4},\,10^{-3},\,6.25\times 10^{-3}]$, corresponding to values of the cooling parameter $\tau \simeq [0,\,3\times 10^{-3},\,1.7\times 10^{-1},\,4.0\times 10^{-1},\,4.0,\,2.4\times 10^{3}]$ at the snapshot time. X-mode emission is quenched for $\tau \gg 1$ and $\Ma < \Ma^{\rm crit}$, as shown in Tab~\ref{tab:sims}, corresponding to $\Ma/\Ma^{\rm crit}\simeq \left[\,20,\,10, \,3,\,0.8,\,0.5,\,0.2\right]$. The shock transition is centered on $\wpp x/c\,=\,0$. $\mathcal{W}\,=\,\int_{\rm prec} B_0^2\,{\rm d}x$.
        }
		\label{fig:EM_sh}
	\end{center}
\end{figure*}

The effect on shock cooling has been investigated by~\citet {Zhang_2025} independently. No clear scaling between shock parameters and efficiency is found in their study. We here discuss their result in light of~\citet{beloborodov_monster_2023} and our analysis. Performing a series of simulations with $\sigma_{\rm u}=10$, they observe a dependence of the efficiency $f_\xi$ on the local density, modifying the cooling strength. By normalizing all simulations in terms of the cooling parameter $\tau$, we have that the density sequences collapse onto a single function of $\tau$ as shown in Fig.~\ref{fig:zhang}. The same figure also confirms the scaling of the shock jump condition, 
$ \gamma_{\rm sh|d}/\sqrt{\sigma_{\rm u}}\propto \left( 1 + \tau \right)^{1/7}$,  shown in the top panel of Fig.~\ref{fig:MAcrit}. The efficiency peaks at $\tau\,\simeq\,5\times10^{-2}$ and falls by an order of magnitude at $\tau\,\simeq\,10^3$. The cutoff  in efficiency is consistent with the scaling $f_{\xi} \simeq 10^{-3}\,\tau^{-2/7}$ for $\tau \gg 1$. The $\tau^{-2/7}$ cutoff is kinematic rather than dynamical, and follows from the increased shock velocity at large $\tau$. Given $\xi_B = \delta B^2/B_0^2$ and the jump conditions derived in~\citet{gallant_relativistic_1992}, the efficiency $f_\xi$ scales as~\citep{plotnikov_synchrotron_2019}:
\begin{align} \label{eq:fxi_def}
f_\xi \,&=\, \xi_B \left( \frac{\sigma_{\rm u}}{1+\sigma_{\rm u}} \right) \left( \frac{1-\beta_{\rm s|d}}{\beta_0 + \beta_{\rm s|d}} \right)\,,\nonumber \\
&\simeq\,\frac{\xi_B}{4 \gamma_{\rm sh |d}^2} \, \simeq \,  \frac{\xi_B}{4} \,\sigma_{\rm u}^{-1} \,(1 + \tau)^{-2/7} \,.
\end{align}
The scaling at large $\tau$ is well recovered in PIC simulations, as shown in Fig.~\ref{fig:zhang}, where we have $f_\xi \simeq 10^{-2}\,\sigma_{\rm u}^{-1} \, \tau^{-2/7}$, for $\tau \gg 1$ and $\sigma_{\rm u} \,=\, 10$.

As for the peak frequency, there is good agreement between the result of ~\citet {Zhang_2025} (Fig. 3.c) and our prediction, Eq.~\eqref{eq:peakom}, for which 
\begin{align}
    \omega_{\rm |d}^{\rm peak} \simeq \tfrac{1}{2} c k_{\rm sol|d} \simeq 5 \, \tau^{2/7} \sqrt{\sigma_{\rm u}} \omega_{\rm p} \,,
\end{align}
at $\tau > 1$. The scaling of the peak frequency is also in agreement with the independent derivation obtained in~\citet{Beloborodov_2026}, where the derived peak frequency scales as $\omega_{\rm |d}^{\rm peak} \,\simeq\,3 \,\tau^{2/7} \sqrt{\sigma_{\rm u}} \omega_{\rm p}$ (see Eq. 40 there).

To summarize, radiation reaction controls the coherent, $X$-mode emission of the monster shock through the single cooling parameter $\tau$, which measures the radiative losses over a gyration time. From the multifluid model we obtained a general threshold for coherent emission, $\Ma^{\rm crit} \simeq 2 + 12.6\,\tau^{0.45}$, set by the onset of chaotic beam dynamics; a peak frequency $\omega_{|d} \simeq 5\,\tau^{2/7}\sqrt{\sigma_{\rm u}}\,\omega_{\rm p}$ for $\tau \gg 1$; and, for the emission efficiency, a kinematic cutoff $f_\xi \simeq 10^{-3}\,\tau^{-2/7}$ following from the increased shock velocity at strong cooling. The suite of PIC simulations confirms these predictions: $X$-mode emission is progressively suppressed as $\tau$ increases and $\Ma/\Ma^{\rm crit}$ falls below unity, the peak frequency upshifts as $\tau^{2/7}$, and the spectrum becomes narrow-banded near threshold, a feature we attribute to the transition to the subcritical regime rather than to cooling itself. Recast in terms of $\tau$, the density-dependent efficiencies reported by \citet{Zhang_2025} collapse onto a single curve consistent with our scaling, and the peak frequency agrees with the independent derivation of \citet{Beloborodov_2026}.

\section{Application to monster shocks} \label{sec:monster}

\begin{figure*}[t]
	\begin{center}
		\includegraphics[width=1\textwidth]{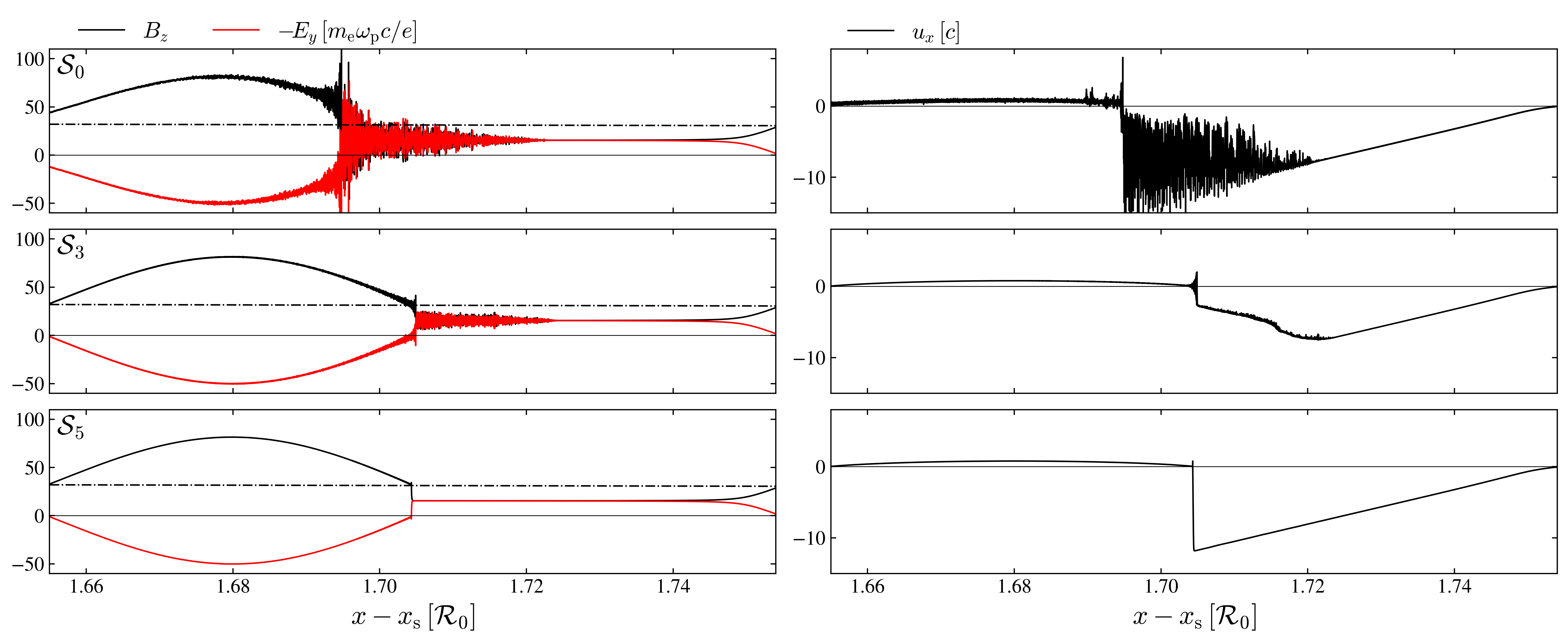}
		\caption{
        Snapshots from the evolution of the FMS wave for simulations
        $\mathcal{S}_0$, $\mathcal{S}_3$, and $\mathcal{S}_5$ (from top to bottom).
        All profiles are shown relative to the steepening point and normalized to the
        gradient length $\mathcal{R}_0$.
        The left panels display the magnetic-field (black) and electric-field (red)
        profiles, while the right panels show the corresponding longitudinal four-velocity
        of the flow. 
        }
		\label{fig:MS_global}
	\end{center}
\end{figure*}

We now apply the above results to the emission of precursor waves in monster shocks, which we associate with a subclass of observed FRBs. 
For a conservative production efficiency of $10^{-3}$, as found in \citep{vanthieghem_fast_2025},
the FRB luminosity is $L_{\rm FRB} = 10^{40} L_{43}$ erg s$^{-1}$, where $L = 10^{43}L_{43}$ erg s$^{-1}$ denotes the luminosity of the steepened kHz wave 
that drives the monster shock.  

It is important to note that the relevant quantity in the context of monster shocks is not the shock efficiency, $f_{\xi}$, but the fraction of the kHz wave energy converted into $X$-modes, $\epsilon_{\rm X}$. In the downstream frame, the incoming energy converted into X-modes during a time interval ${\rm d}t$ is $f_{\xi}\,(1+\sigma_{\rm u})\,\gamma_{\rm u|d}^2\, n_{\rm u}\, \me c^3\,{\rm d}t \simeq f_{\xi}\,\sigma_{\rm u}\,\gamma_{\rm u|d}^2\, n_{\rm u}\, \me c^3\,{\rm d}t$. Expressed in terms of the wavelength increment swept up by the shock, $c\,{\rm d}t = 2\gamma_{\rm sh|d}^{2}\,{\rm d}l$, and using $f_\xi \simeq \xi_{\rm B}/4\gamma_{\rm sh|d}^2$, we obtain
\begin{align}
\epsilon_{\rm X} \,&\simeq\, \frac{\xi_{\rm B}}{2}\,
\frac{\int_0^{\lambda/2} \sigma_{\rm u}\,\gamma_{\rm u|d}^2\, n_{\rm u}\, \me c^2\,{\rm d}l}
{\lambda\,E_{\rm w}}\,, \nonumber \\
&\sim\, \frac{\xi_{\rm B}}{4} \simeq 10^{-3} - 10^{-2}\,,
\end{align}
where $E_{\rm w}$ is the energy density of the kHz wave. The range $\epsilon_{\rm X} \simeq 10^{-3}$-$10^{-2}$ reflects the weak dependence of $\xi_{\rm B}$ on the cooling parameter in the simulations of \citet{Zhang_2025}, recast in terms of $\tau$. The magnetic energy fraction of the precursor increases from $\xi_{\rm B} \simeq 10^{-2}$ for $\tau \lesssim 10^{-5}$ to $\xi_{\rm B} \simeq 5\times10^{-2}$ for $ 1 \lesssim \tau \lesssim 10^3$. Interestingly, the efficiency therefore increases moderately with cooling, so that strongly cooled monster shocks lie at the upper end of this range, $\epsilon_{\rm X} \sim 10^{-2}$.
Because the downstream plasma is nearly at rest in the star frame, downstream- and star-frame quantities coincide to within factors of order $\gamma_{\rm d|\star}$. The dependence on $\gamma_{\rm sh|d}$, and hence on $\tau$, cancels, so that $\epsilon_{\rm X}$ is determined by $\xi_{\rm B}$, whereas the time required to dissipate the wave, and hence the extent of the dissipation zone, increases as $\gamma_{\rm sh|d}^2 \propto \sigma_{\rm u} \tau^{2/7}$.

We suppose that the kHz wave propagates in a dipolar magnetar magnetosphere with an equatorial magnetic field $B\,=\,10^{15}\,B_{15}\,(R/r)^3\,{\rm G}$, where $R\,=\,10^6\,{\rm cm}$ is the neutron star radius.
The shock forms at a radius $r_{\rm s}\,=\,2\times10^8\,B_{15}^{1/2}\,L_{43}^{-1/4}\,{\rm cm}$, and, when radiative cooling is neglected, dissipates the lower half of the kHz wave over the interval $r_{\rm s}\lesssim r\lesssim3r_{\rm s}$ \citep{beloborodov_monster_2023}. The modified jump conditions alter the extent of the dissipation zone. In the strong-cooling regime, if $\tau$ remained constant, the expected interval would instead be $r_{\rm s}\lesssim r\lesssim3\tau^{2/7}r_{\rm s}$. As we show below, however, $\tau$ decreases rapidly with $r/r_{\rm s}$, and the cooling-free estimate remains a good approximation.

The Lorentz factor of the upstream plasma reaches its maximum shortly after shock formation and subsequently decreases as $\gamma_{\rm u}\propto\pi(r_{\rm s}/r)^4$ as the shock strengthens (Eq.~75 of \citealt{beloborodov_monster_2023}; see also \citealt{Bernardi2025}). The upstream magnetization in the dissipation zone is therefore
$\sigma_{\rm u} = \sigma_{\rm bg}/2\gamma_{\rm u} \approx \omega\, r/(2 c)$. It attains its minimum value at $r=r_{\rm s}$:
 $\sigma_{\rm u,min} \approx 2\times 10^2\,\,B_{15}^{1/2}\,\nu_4\,L_{43}^{-1/4}$, corresponding to $\sigma_{\rm u,min}\simeq2 \times 10^2$ for our fiducial parameters.

Since the shock Lorentz factor is $\gamma_{\rm sh}=2\gamma_{\rm u}\gamma_{\rm sh|u}$, with $\gamma_{\rm sh|u}\simeq\sqrt{\sigma_{\rm u}}$ when the $\tau^{1/7}$ correction is neglected, the corresponding Alfvénic Mach number scales as $\Ma = \gamma_{\rm sh}/\sqrt{\sigma_{\rm u}} \sim \gamma_{\rm u}$.
Thus, $\Ma$ inherits the $(r_{\rm s}/r)^4$ decline of the upstream Lorentz factor, rather than steepening with the magnetization. The proper density immediately upstream of the shock is $n_{\rm u}=n_{\rm bg}/(2\gamma_{\rm u})$, where $n_{\rm bg}\propto(r_{\rm s}/r)^3$ is the background pair density in the magnetosphere.
In terms of the magnetar rotation period $P$ (measured in seconds), and the kHz wave frequency $\nu = 10^4 \nu_4$ Hz, we find the following scaling relations:
\begin{align}
    &\gamma_{\rm u} \,\sim\, 4\times 10^5\, \varpi \left(\frac{r_{\rm s}}{r}\right)^4 \, B_{15}^{-1}\,\mathcal{M}_6^{-1}\, P_0 \, \nu_4^{-1}\,L_{43}    \label{eq:scal1} \,,\\ 
    &M_{\rm A}\,\sim\, 6\times 10^5 \, \varpi \left(\frac{r_{\rm s}}{r}\right)^4\,B_{15}^{-1}\,\mathcal{M}_6^{-1}\, P_0 \, \nu_4^{-1}\, L_{43}\,,\\
    &\tau \,\sim\,3 \times 10^3 \, \varpi^{2} \left(\frac{r_{\rm s}}{r}\right)^{11} \, B_{15}^{-5/2}\,\mathcal{M}_6^{-2}\, P_0^2\,\nu_4^{-2}\,L_{43}^{11/4}\,, \\
    & \frac{\Ma}{\Ma^{\rm crit}} \,\sim\,10^3 \, \varpi^{0.1} \left(\frac{r_{\rm s}}{r}\right)^{-0.95}  \, \nonumber\\ & \qquad \qquad \qquad \times  B_{15}^{1/8}\,\mathcal{M}_6^{-0.1}\, P_0^{0.1}\,\nu_4^{-0.1}\,L_{43}^{- 0.24} \label{eq:M_crt/M} \,, 
\end{align} 
where we adopt a pair multiplicity $\mathcal{M}=10^6\,\mathcal{M}_6$ with respect to the Goldreich-Julian density, and where $\varpi = 1$ for $r \lesssim 2\,r_{\rm s}$ and $\varpi=\pi$ elsewhere.  In the last equation, $\Ma^{\rm crit}$ is the threshold Mach number for X-mode generation, Eq. \eqref{eq:M_A_crit}. 

Note that $\tau(r)\propto(r_{\rm s}/r)^{11}$ throughout the dissipation zone, $r_{\rm s}\lesssim r\lesssim3r_{\rm s}$. For these parameters, the monster shock forms in the strong-cooling regime, although the importance of cooling decreases rapidly as the shock propagates outward. 
Importantly, Eq.~\eqref{eq:M_crt/M} shows that monster shocks remain robustly above the threshold for X-mode generation throughout the parameter range inferred from observations.

We now scale the frequency of the precursor wave to a realistic magnetar.  The 
background plasma frequency may be written as
\begin{equation}
    \omega_{\rm p, bg}\approx 10^2 B_{15}^{-1/4} L_{43}^{3/8} {\cal M}_6^{1/2} P_0^{-1/2}(r_s/r)^{3/2}  \, \, \text{GHz}.
 \end{equation}
 At the end of the dissipation zone, $r\approx 3r_s$, the corresponding plasma frequency is $\nu_{\rm p} = \omega_{\rm p}/2\pi \sim 3 \, {\cal M}_6^{1/2}$ GHz, for our fiducial normalization. 
 For the Galactic FRB we then estimate $\nu_{\rm p}\approx 0.5 \mathcal{M}^{1/2}_6$ GHz at the onset of the precursor wave escape zone.
 The proper plasma frequency just upstream of the shock is $\omega_{\rm p} =\omega_{\rm p, bg}/\sqrt{2 \gamma_{\rm u}}$. In the weak cooling regime the 
 characteristic frequency of the precursor wave is $\omega_{\rm | d}^{\rm peak} \approx 2\sqrt{\sigma_{\rm u}} \omega_{\rm p} = 2\sqrt{\sigma_{\rm u}/\gamma_{\rm u}} \, \omega_{\rm p, bg}$. For $\nu_\mathrm{FRB} \simeq \omega_{\rm | d}^{\rm peak}/2\pi$ we have in the regime of weak cooling, $\tau \ll 1$:
\begin{align}
\nu_\mathrm{FRB}  &\sim 1 \,\varpi^{-1}\, \left(\frac{r_{\rm s}}{r}\right)^{-1} \quad \text {GHz}  \nonumber \\
&  \quad \times \,B_{15}^{1/2} \mathcal{M}_6 P_0^{-1} \nu_4 L_{43}^{-1/4} \label{eq:nu_tau_0} \quad   \,.
\end{align}
In the regime of strong cooling, $\tau \gg1$, we have:
\begin{align}
\nu_\mathrm{FRB} \, &\sim 20 \,\varpi^{-3/7}\, \left(\frac{r_{\rm s}}{r}\right)^{15/7} \quad \text {GHz}  \nonumber \\
&  \quad \times \,B_{15}^{-3/14} \mathcal{M}_6^{3/7} P_0^{-3/7} \nu_4^{3/7} L_{43}^{15/28} \label{eq:nu_tau}  \,.
\end{align}
Comparison with observations suggests ${\cal M}_6\sim10^{-2}$, consistent with the pair multiplicity expected to be generated in twisted magnetic flux tubes in magnetar magnetospheres \citep{mahlmann2026}.

In the strong-cooling regime, the precursor-wave frequency is significantly upshifted, as shown by Eq.~\eqref{eq:nu_tau}. This trend is evident in the right panel of Fig.~\ref{fig:EM_sh}, where the spectral peak shifts to progressively larger values of $k$ as $\tau$ increases. Consequently, bright bursts, for which radiative cooling is important, are likely to remain in the kinetic regime as they escape, avoiding nonlinear decay \citep{golbraikh2023,solanki2026}.

\begin{figure}[t]
	\begin{center}
		\includegraphics[width=1\columnwidth]{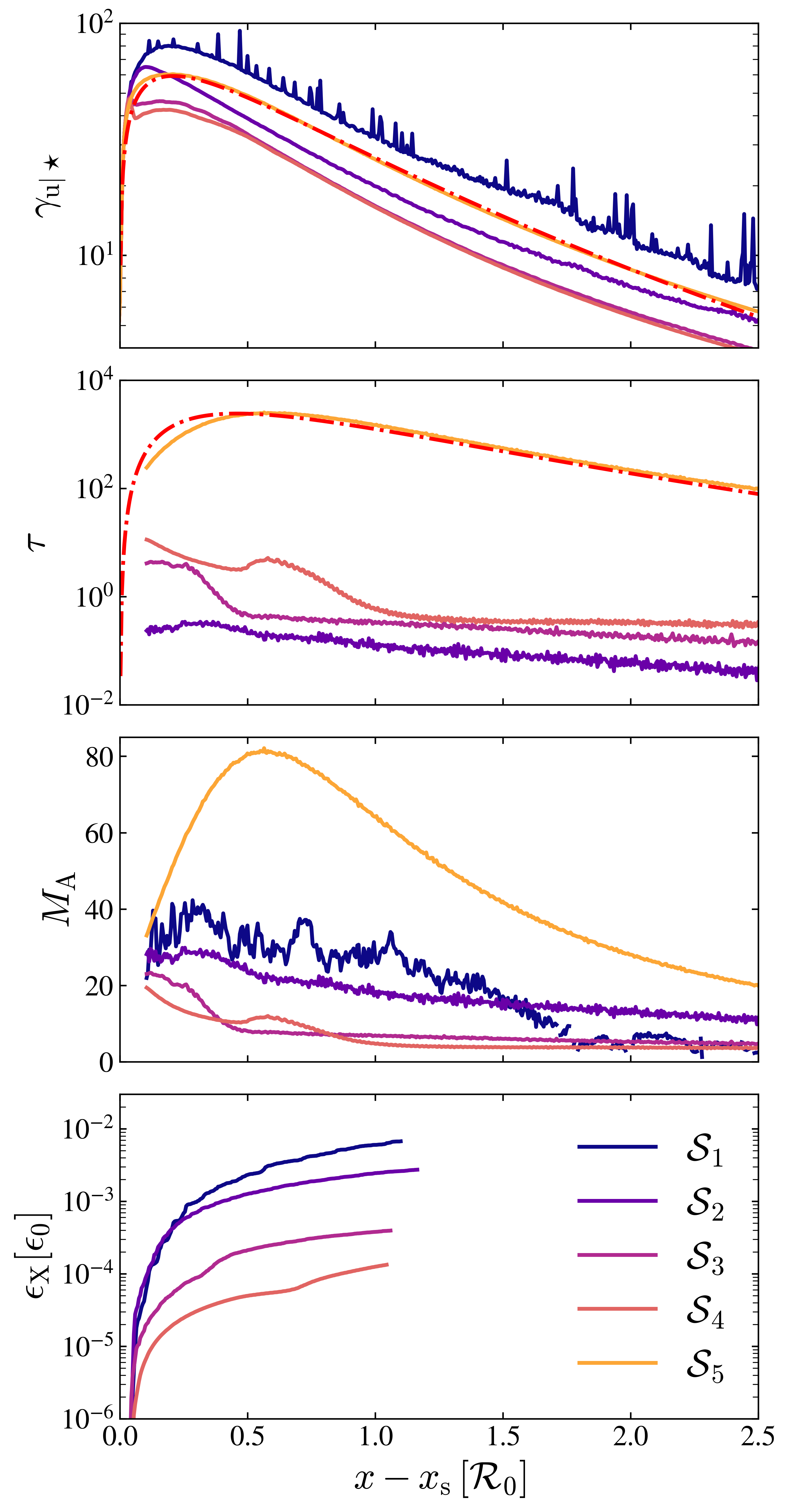}
		\caption{
        Evolution of shock and radiative quantities for simulations $\mathcal{S}_1$–$\mathcal{S}_5$. From top to bottom: $\gamma_{\rm u|\star}$, $\tau$, $M_{\rm A}$, and the fraction of energy in X-modes, $\epsilon_{\rm X}$, normalized to the incoming wave energy, $\epsilon_0$. All quantities are shown as functions of the shock position relative to the steepening point $x - x_{\rm s}$ in units of $\mathcal{R}_0$. The dot-dashed red curves correspond to the analytical estimates of~\citet{beloborodov_monster_2023} (Eqs. (75) - (94)).
        }
		\label{fig:MS_prof}
	\end{center}
\end{figure}

\subsection{Interaction with the upstream flow \label{subsec:upstream}}

For typical FRB luminosities, the precursor wave is relativistically strong, with a strength parameter $a_0 = e E/m_ec\omega \gg 1$. 
Here $E = (2 L_{\rm FRB}/c r^2)^{1/2}$.
For the production efficiency of $10^{-3}$ found in \citet{vanthieghem_fast_2025} for weak cooling, and precursor wave frequency of $\nu = 10^9 \nu_9$ Hz, we have:
\begin{equation}
a_0 \approx 10^4 L_{43}^{3/4} B_{15}^{-1/2} \nu_9^{-1} (r_s/r) \label{eq:ao}.
\end{equation}
Our present analysis indicates that a comparable efficiency is maintained in strongly cooled, supercritical shocks, so Eq.~\eqref{eq:ao} remains applicable in this regime. Combining Eqs.~\eqref{eq:scal1} and \eqref{eq:ao} gives:
\begin{equation}
\frac{\gamma_{\rm u}}{a_0} \approx 40 \,\varpi \, B_{15}^{-1/2} L_{43}^{1/4} ({\cal M}_6 \nu_4)^{-1} \, P_0\,\nu_9 \, (r_s/r)^{3}.
\label{eq:gamma/a}
\end{equation}

When the frequency of the precursor wave lies in the interval $\omega'_B/a_0 < \omega' < a_0\omega'_B$, as measured in the rest frame of the upstream fluid,
the plasma inside the wave may be heated through synchrotron absorption \citep{lyubarsky2018,beloborodov2022,sobacchi2024}.  This might significantly reduce the efficiency of the precursor wave emission. 
At higher frequencies, $\omega' > a_0 \omega'_B$, synchrotron absorption is strongly suppressed \citep{lyubarsky2018,sobacchi2024,sobacchi2025} and we do not expect the efficiency to be affected. 

The frequency of the precursor wave is $\omega \sim \zeta \sqrt{\sigma_{\rm u}} \, \omega_{\rm p} = \zeta \, \omega_B'$ (Eqs. \ref{eq:nu_tau_0}, \ref{eq:nu_tau}), where $\zeta = 2$ in the weak cooling regime ($\tau\ll1$)
and $\zeta = 5\tau^{2/7}$ in the strong cooling regime ($\tau\gg1$), and the prime denotes
quantities in the rest frame of the upstream fluid. The dispersion relation in the Lab
frame is approximately $k^2 = \omega^2 - \omega^{2}_{\rm p}$ \citep{plotnikov_synchrotron_2019,vanthieghem_fast_2025}.
Transforming to the fluid frame, we obtain:
\begin{align} \label{eq:omegap}
    \omega' \,&=\, [\sqrt{\zeta^2 \sigma_{\rm u} - 1}  + \zeta \sqrt{\sigma_{\rm u}}]\gamma_{\rm u} \omega_{\rm p} \nonumber \\
    &\simeq \,2 \gamma_{\rm u} \omega \,=\,\frac{2\zeta\gamma_{\rm u}}{a_0} a_0\omega'_B. 
\end{align}
From Eq. \eqref{eq:gamma/a} we conclude that $\omega' \gg a_0 \omega'_B$ in the dissipation zone; hence, significant heating by the wave is not expected within the dissipation region \citep{sobacchi2024,sobacchi2025}.

However, the question of whether the precursor wave can escape once dissipation at the shock is complete remains open. Interestingly, Eqs.~\eqref{eq:gamma/a} and \eqref{eq:omegap} imply that $\omega' \sim a_0\omega_B'$ at $r \sim 7\,r_{\rm s}$, suggesting that the subsequent evolution of the wave may involve interesting dynamics. 
We stress, however, that even a small modulation of the upstream magnetic field, with an amplitude comparable to that of the precursor wave, could allow the wave to escape by riding on the propagating pulse, provided that the modulation wavelength is much longer than that of the precursor wave.

The precursor wave can also modify the flow ahead of the shock. Interestingly, this coupling is already evident in our simulations, even though the precursor remains in the linear regime, $a_0 \ll 1$. As shown in Fig.~\ref{fig:MS_global}, the upstream flow in $\mathcal{S}_3$ is decelerated throughout the region occupied by the precursor wave. By contrast, in the uncooled run $\mathcal{S}_0$, the wave merely modulates the incoming flow without producing a net deceleration, while in the strongly cooled run $\mathcal{S}_5$, where the emission is quenched, the flow reaches the shock unperturbed. Consequently, the upstream Lorentz factors in $\mathcal{S}_3$ and $\mathcal{S}_4$ remain smaller than the analytical estimate ~\cite{beloborodov_monster_2023} by a factor of $\approx 1.5$ throughout the dissipation zone, whereas $\mathcal{S}_5$ closely follows this estimate (Fig.~\ref{fig:MS_prof}). In Appendix~\ref{app:coupling}, we show that this deceleration results from radiative damping of the precursor wave, which transfers its momentum to the incoming plasma. The effect is therefore strongest at intermediate cooling strengths, for which the wave is still emitted but is efficiently absorbed ahead of the shock. In monster shocks, where $a_0 \gg 1$, the coupling is nonlinear but may produce a similar deceleration~\citep{Beloborodov_2026}. If the upstream Lorentz factor is reduced by a factor $\eta = \hat{\gamma}_{\rm u}/\gamma_{\rm u}$, the cooling parameter decreases as $\eta^{2}$ and the peak frequency increases as $\eta^{-3/7}$, whereas supercriticality with respect to the emission threshold, $\Ma/\Ma^{\rm crit}$, scales only as $\eta^{1/10}$ (Eq.~\eqref{eq:eta_scal}). The values of $\gamma_{\rm u}$, $\Ma$, and $\tau$ given by Eq.~\eqref{eq:scal1} should therefore be regarded as upper bounds. Even for $\eta = 0.1$, however, monster shocks remain far above the emission threshold.

In summary, monster shocks are strongly cooled at the base of the dissipation zone, where $\tau \gg 1$, although the cooling parameter declines steeply with radius, $\tau \propto (r_{\rm s}/r)^{11}$. Nevertheless, they remain far above the emission threshold throughout the dissipation zone, with $\Ma/\Ma^{\rm crit} \sim 10^3$. Radiation reaction therefore does not quench the precursor emission, and the FRB luminosity is set by the luminosity of the driving kHz wave and a production efficiency of $\sim 10^{-3}$. Instead, cooling shifts the peak frequency upwards, as $\tau^{2/7}$, and pushes the escaping wave deeper into the kinetic regime. The precursor wave may also decelerate the incoming plasma, thereby reducing the cooling parameter while having little effect on the ratio $\Ma/\Ma^{\rm crit}$. Within the dissipation zone, $\omega' \gg a_0\,\omega'_B$, so the wave is not expected to heat the upstream plasma through synchrotron absorption. Whether it can escape beyond this zone remains an open question.

\section{Conclusion \label{sec:conclusion}}

We have investigated the effects of radiation reaction on the structure of relativistically magnetized pair shocks and on the coherent precursor emission they generate, combining a generalized stationary multifluid model with fully kinetic simulations performed using the \entity~code \citep{Hakobyan2025,Boss_2026}. Specifically, we extended the multifluid model of \citet{vanthieghem_relativistically_2025}, in which the plasma is represented by $2N$ cold beams coupled through the electromagnetic field, by incorporating the reduced Landau--Lifshitz force, given by Eq.~\eqref{eq:RRforce}, into the beam equations of motion.

We find that the shock transition is governed by only two dimensionless parameters: the Alfvénic Mach number $\Ma$ and the cooling parameter $\tau$, defined in Eq.~\eqref{eq:tau}, which measures the radiative energy losses over a gyration period. Increasing $\tau$ reduces the amplitude of the solitary structure, while enhancing its coherence and shortening its wavelength. The competition among these effects leads to a non-monotonic dependence of the precursor emissivity on cooling strength. We have shown that chaotic beam dynamics, which is necessary for the formation of the ring distributions that drive the synchrotron maser, develops only above a critical Mach number,
$\Ma > \Ma^{\rm crit} \simeq 2 + 12.6\,\tau^{0.45}$; below this threshold, phase mixing is negligible and no significant coherent emission is expected. Above it, the peak frequency is determined by the spacing between successive solitons and, in the strong-cooling regime, scales as $\omega_{\rm |d}^{\rm peak} \simeq 5\,\tau^{2/7} \sqrt{\sigma_{\rm u}}\,\omega_{\rm p}$ for $\tau \gg 1$.

A suite of kinetic simulations confirms these predictions over five decades in cooling strength. The $X$-mode emission is progressively suppressed as $\tau$ increases and $\Ma/\Ma^{\rm crit}$ falls below unity; the peak frequency increases as $\tau^{2/7}$, and the spectrum becomes strongly narrow-banded near the threshold. Because the same spectral narrowing is observed in an uncooled shock at $\Ma\simeq2$, we attribute it to the transition into the subcritical regime rather than to cooling itself. We therefore expect narrow-banding to be a generic signature of shocks operating close to criticality. When recast in terms of $\tau$, the density-dependent efficiencies reported by \citet{Zhang_2025} collapse onto a single function of the cooling parameter. Moreover, the residual decline at large $\tau$, $f_\xi \simeq 10^{-2}\, \sigma_{\rm u}^{-1}\,\tau^{-2/7}$,
is kinematic rather than dynamical in origin. It follows that the fraction of energy tapped into $X$-modes, when integrated over the whole dissipation region, is only a function of an almost constant $\xi_{B}\,\sim 10^{-2} - 10^{-1}$~\cite{Zhang_2025}. The fraction of the available wave energy converted into $X$-mode emission is therefore only weakly affected by cooling and maximized at large $\tau$.

Applied to monster shocks in magnetar magnetospheres, these results lead to a simple picture. The cooling parameter is large at the base of the dissipation zone $r_s$ but declines steeply with radius, $\tau \propto (r_{\rm s}/r)^{11}$. 
By contrast, the Mach number remains far above the threshold for $X$-mode generation throughout the dissipation zone, with
$\Ma/\Ma^{\rm crit} \sim 10^3 \,(r_{\rm s}/r)^{-0.95}$.
Monster shocks are therefore both strongly cooled and strongly supercritical for any reasonable choice of magnetospheric parameters. Radiation reaction consequently does not quench the precursor emission, regardless of the luminosity of the driving kHz wave: the luminosity of the resulting burst is determined by the luminosity of the kHz wave and by the production efficiency $\epsilon_{X} \sim 10^{-3} - 10^{-2}$, close to the value found by \citet{vanthieghem_fast_2025}. Strong cooling leads to $\epsilon_{X} \sim 10^{-2}$ and alters the spectrum and geometry of the emitting region. It shifts the spectral peak to higher frequencies, pushes the escaping wave deeper into the kinetic regime, and extends the dissipation zone. Bright bursts, for which cooling is most important, should therefore be emitted at the highest frequencies.

Several aspects of the problem remain open. The scaling relations derived here assume weak coupling between the shock and the precursor wave. Whether the wave can escape after dissipation at the shock is complete is a separate question that cannot be addressed by our present setup, which follows the shock only across the dissipation zone. The subsequent evolution of the wave will be investigated in a forthcoming study. The physical mechanism underlying the coherent $X$-mode emission also remains to be clarified and will be addressed in future work building on \citet{vanthieghem_relativistically_2025}. Finally, the simulations presented here are one-dimensional; the multidimensional dynamics of the soliton train and the resulting angular structure of the precursor emission are left for future investigation.

\begin{acknowledgments}
AV is supported by Université PSL through the PSL Young Researcher Starting Grant 2025 (2025-396) and received funding under the French Government program “Investissements d’Avenir."  AL acknowledges support by a grant from the Simons Foundation (MP-SCMPS-00001470).
This project was provided with computing HPC and storage resources by GENCI at TGCC and CINES, thanks to the grants 2025-A0180415130 and 2026-A0200415130 on the supercomputer Joliot Curie and Adastra's ROME, V100, and MI250x partitions.
\end{acknowledgments}

\bibliography{Bib2}

@article{vanthieghem_fast_2025,
	title = {Fast {Radio} {Bursts} as {Precursor} {Radio} {Emission} from {Monster} {Shocks}},
	volume = {134},
	url = {https://link.aps.org/doi/10.1103/PhysRevLett.134.035201},
	doi = {10.1103/PhysRevLett.134.035201},
	number = {3},
	urldate = {2025-05-14},
	journal = {Phys. Rev. Lett.},
	author = {Vanthieghem, A. and Levinson, A.},
	month = jan,
	year = {2025},
	note = {Publisher: American Physical Society},
	pages = {035201},
}

@ARTICLE{Zhang_2025,
       author = {{Zhang}, Yu and {Yang}, Yuanpei and {Ji}, Liangliang},
        title = "{Radiation Reaction effects on Coherent Emission in Relativistic Magnetized Shocks}",
      journal = {arXiv e-prints},
         year = 2025,
        month = feb,
          eid = {arXiv:2502.14550},
        pages = {arXiv:2502.14550},
          doi = {10.48550/arXiv.2502.14550},
archivePrefix = {arXiv},
       eprint = {2502.14550},
 primaryClass = {astro-ph.HE},
       adsurl = {https://ui.adsabs.harvard.edu/abs/2025arXiv250214550Z}
}

@ARTICLE{Vay_2020,
       author = {{Vay}, J.~L.},
        title = "{Simulation of plasma accelerators with the Particle-In-Cell method}",
      journal = {arXiv e-prints},
         year = 2020,
        month = aug,
          eid = {arXiv:2008.07300},
        pages = {arXiv:2008.07300},
          doi = {10.48550/arXiv.2008.07300},
archivePrefix = {arXiv},
       eprint = {2008.07300},
 primaryClass = {physics.comp-ph},
       adsurl = {https://ui.adsabs.harvard.edu/abs/2020arXiv200807300V}
}

@ARTICLE{Bilbao_2025,
       author = {{Bilbao}, Pablo and {Silva}, Thales and {Silva}, Lu{\'\i}s O.},
        title = "{Radiative cooling induced coherent maser emission in relativistic plasmas}",
      journal = {Science Advances},
         year = 2025,
        month = apr,
       volume = {11},
       number = {15},
          eid = {eadt8912},
        pages = {eadt8912},
          doi = {10.1126/sciadv.adt8912},
archivePrefix = {arXiv},
       eprint = {2409.18955},
 primaryClass = {physics.plasm-ph},
       adsurl = {https://ui.adsabs.harvard.edu/abs/2025SciA...11.8912B}
}

@article{vanthieghem_relativistically_2025,
	title = {Relativistically magnetized collisionless shocks in pair plasma: {Solitons}, chaos, and thermalization},
	volume = {111},
	shorttitle = {Relativistically magnetized collisionless shocks in pair plasma},
	url = {https://link.aps.org/doi/10.1103/PhysRevE.111.045209},
	doi = {10.1103/PhysRevE.111.045209},
	number = {4},
	urldate = {2025-05-14},
	journal = {Phys. Rev. E},
	author = {Vanthieghem, Arno and Levinson, Amir},
	month = apr,
	year = {2025},
	note = {Publisher: American Physical Society},
	pages = {045209},
}

@article{beloborodov_monster_2023,
	title = {Monster {Radiative} {Shocks} in the {Perturbed} {Magnetospheres} of {Neutron} {Stars}},
	volume = {959},
	issn = {0004-637X},
	url = {https://dx.doi.org/10.3847/1538-4357/acf659},
	doi = {10.3847/1538-4357/acf659},
	language = {en},
	number = {1},
	urldate = {2025-05-14},
	journal = {ApJ},
	author = {Beloborodov, Andrei M.},
	month = dec,
	year = {2023},
	note = {Publisher: The American Astronomical Society},
	pages = {34},
}

@ARTICLE{Beloborodov_2026,
       author = {{Beloborodov}, Andrei M.},
        title = "{Radio precursors of monster shocks: a mechanism for fast radio bursts from SGR 1935+2154}",
      journal = {arXiv e-prints},
         year = 2026,
        month = jun,
          eid = {arXiv:2606.10189},
        pages = {arXiv:2606.10189},
          doi = {10.48550/arXiv.2606.10189},
archivePrefix = {arXiv},
       eprint = {2606.10189},
 primaryClass = {astro-ph.HE},
       adsurl = {https://ui.adsabs.harvard.edu/abs/2026arXiv260610189B}
}

@article{waxman_origin_2017,
	title = {On the {Origin} of {Fast} {Radio} {Bursts} ({FRBs})},
	volume = {842},
	issn = {0004-637X},
	url = {https://dx.doi.org/10.3847/1538-4357/aa713e},
	doi = {10.3847/1538-4357/aa713e},
	language = {en},
	number = {1},
	urldate = {2025-05-14},
	journal = {ApJ},
	author = {Waxman, Eli},
	month = jun,
	year = {2017},
	note = {Publisher: The American Astronomical Society},
	pages = {34},
}

@article{lyubarsky_model_2014,
	title = {A model for fast extragalactic radio bursts},
	volume = {442},
	issn = {1745-3925},
	url = {https://doi.org/10.1093/mnrasl/slu046},
	doi = {10.1093/mnrasl/slu046},
	number = {1},
	urldate = {2025-05-14},
	journal = {Monthly Notices of the Royal Astronomical Society: Letters},
	author = {Lyubarsky, Yuri},
	month = jul,
	year = {2014},
	pages = {L9--L13},
}

@article{lyubarsky_emission_2021,
	title = {Emission {Mechanisms} of {Fast} {Radio} {Bursts}},
	volume = {7},
	copyright = {http://creativecommons.org/licenses/by/3.0/},
	issn = {2218-1997},
	url = {https://www.mdpi.com/2218-1997/7/3/56},
	doi = {10.3390/universe7030056},
	language = {en},
	number = {3},
	urldate = {2025-05-14},
	journal = {Universe},
	author = {Lyubarsky, Yuri},
	month = mar,
	year = {2021},
	note = {Number: 3
Publisher: Multidisciplinary Digital Publishing Institute},
	pages = {56},
}

@article{metzger_fast_2019,
	title = {Fast radio bursts as synchrotron maser emission from decelerating relativistic blast waves},
	volume = {485},
	issn = {0035-8711},
	url = {https://doi.org/10.1093/mnras/stz700},
	doi = {10.1093/mnras/stz700},
	number = {3},
	urldate = {2025-05-14},
	journal = {Monthly Notices of the Royal Astronomical Society},
	author = {Metzger, Brian D and Margalit, Ben and Sironi, Lorenzo},
	month = may,
	year = {2019},
	pages = {4091--4106},
}

@misc{chen_propagation_2022,
	title = {Propagation of a {Strong} {Fast} {Magnetosonic} {Wave} in the {Magnetosphere} of a {Neutron} {Star}},
	copyright = {Creative Commons Attribution 4.0 International},
	url = {https://arxiv.org/abs/2210.13506},
	doi = {10.48550/ARXIV.2210.13506},
	urldate = {2025-05-14},
	publisher = {arXiv},
	author = {Chen, Alexander Y. and Yuan, Yajie and Li, Xinyu and Mahlmann, Jens F.},
	year = {2022},
	note = {Version Number: 1},
}

@article{gallant_relativistic_1992,
	title = {Relativistic, perpendicular shocks in electron-positron plasmas},
	volume = {391},
	issn = {0004-637X, 1538-4357},
	url = {http://adsabs.harvard.edu/doi/10.1086/171326},
	doi = {10.1086/171326},
	language = {en},
	urldate = {2025-05-14},
	journal = {ApJ},
	author = {Gallant, Yves A. and Hoshino, Masahiro and Langdon, A. B. and Arons, Jonathan and Max, Claire E.},
	month = may,
	year = {1992},
	pages = {73},
}

@article{hoshino_preferential_1991,
	title = {Preferential positron heating and acceleration by synchrotron maser instabilities in relativistic positron–electron–proton plasmas},
	volume = {3},
	issn = {0899-8221},
	url = {https://pubs.aip.org/pfb/article/3/3/818/941331/Preferential-positron-heating-and-acceleration-by},
	doi = {10.1063/1.859877},
	language = {en},
	number = {3},
	urldate = {2025-05-14},
	journal = {Physics of Fluids B: Plasma Physics},
	author = {Hoshino, Masahiro and Arons, Jonathan},
	month = mar,
	year = {1991},
	pages = {818--833},
}

@article{iwamoto_precursor_2019,
	title = {Precursor {Wave} {Amplification} by {Ion}–{Electron} {Coupling} through {Wakefield} in {Relativistic} {Shocks}},
	volume = {883},
	issn = {2041-8205, 2041-8213},
	url = {https://iopscience.iop.org/article/10.3847/2041-8213/ab4265},
	doi = {10.3847/2041-8213/ab4265},
	number = {2},
	urldate = {2025-05-14},
	journal = {ApJL},
	author = {Iwamoto, Masanori and Amano, Takanobu and Hoshino, Masahiro and Matsumoto, Yosuke and Niemiec, Jacek and Ligorini, Arianna and Kobzar, Oleh and Pohl, Martin},
	month = oct,
	year = {2019},
	pages = {L35},
}

@article{plotnikov_synchrotron_2019,
	title = {The synchrotron maser emission from relativistic shocks in {Fast} {Radio} {Bursts}: {1D} {PIC} simulations of cold pair plasmas},
	volume = {485},
	copyright = {https://academic.oup.com/journals/pages/open\_access/funder\_policies/chorus/standard\_publication\_model},
	issn = {0035-8711, 1365-2966},
	shorttitle = {The synchrotron maser emission from relativistic shocks in {Fast} {Radio} {Bursts}},
	url = {https://academic.oup.com/mnras/article/485/3/3816/5370092},
	doi = {10.1093/mnras/stz640},
	language = {en},
	number = {3},
	urldate = {2025-05-14},
	journal = {Monthly Notices of the Royal Astronomical Society},
	author = {Plotnikov, Illya and Sironi, Lorenzo},
	month = may,
	year = {2019},
	pages = {3816--3833},
}

@article{sironi_coherent_2021,
	title = {Coherent {Electromagnetic} {Emission} from {Relativistic} {Magnetized} {Shocks}},
	volume = {127},
	issn = {0031-9007, 1079-7114},
	url = {https://link.aps.org/doi/10.1103/PhysRevLett.127.035101},
	doi = {10.1103/PhysRevLett.127.035101},
	language = {en},
	number = {3},
	urldate = {2025-05-14},
	journal = {Phys. Rev. Lett.},
	author = {Sironi, Lorenzo and Plotnikov, Illya and Nättilä, Joonas and Beloborodov, Andrei M.},
	month = jul,
	year = {2021},
	pages = {035101},
}

@ARTICLE{Boss_2026,
       author = {{B{\"o}ss}, Ludwig M. and {Vanthieghem}, Arno and {Hakobyan}, Hayk and {Gorbunov}, Evgeny A. and {Caprioli}, Damiano},
        title = "{Entity -- Hardware-agnostic Particle-in-Cell Code for Plasma Astrophysics. III: Higher-order shape functions \& generalized field stencils}",
      journal = {arXiv e-prints},
         year = 2026,
        month = may,
          eid = {arXiv:2605.15260},
        pages = {arXiv:2605.15260},
          doi = {10.48550/arXiv.2605.15260},
archivePrefix = {arXiv},
       eprint = {2605.15260},
 primaryClass = {astro-ph.HE},
       adsurl = {https://ui.adsabs.harvard.edu/abs/2026arXiv260515260B}
}

@ARTICLE{lyubarsky2018,
       author = {{Lyubarsky}, Yuri},
        title = "{Interaction of the electromagnetic precursor from a relativistic shock with the upstream flow - I. Synchrotron absorption of strong electromagnetic waves}",
      journal = {\mnras},
         year = 2018,
        month = feb,
       volume = {474},
       number = {1},
        pages = {1135-1142},
          doi = {10.1093/mnras/stx2832},
archivePrefix = {arXiv},
       eprint = {1711.09581},
 primaryClass = {astro-ph.HE},
       adsurl = {https://ui.adsabs.harvard.edu/abs/2018MNRAS.474.1135L}
}

@ARTICLE{sobacchi2024,
       author = {{Sobacchi}, E. and {Iwamoto}, M. and {Sironi}, L. and {Piran}, T.},
        title = "{Escape of fast radio bursts from magnetars}",
      journal = {\aap},
         year = 2024,
        month = oct,
       volume = {690},
          eid = {A332},
        pages = {A332},
          doi = {10.1051/0004-6361/202451725},
archivePrefix = {arXiv},
       eprint = {2409.10732},
 primaryClass = {astro-ph.HE},
       adsurl = {https://ui.adsabs.harvard.edu/abs/2024A&A...690A.332S}
}

@ARTICLE{beloborodov2022,
       author = {{Beloborodov}, Andrei M.},
        title = "{Scattering of Ultrastrong Electromagnetic Waves by Magnetized Particles}",
      journal = {\prl},
         year = 2022,
        month = jun,
       volume = {128},
       number = {25},
          eid = {255003},
        pages = {255003},
          doi = {10.1103/PhysRevLett.128.255003},
archivePrefix = {arXiv},
       eprint = {2108.05464},
 primaryClass = {astro-ph.HE},
       adsurl = {https://ui.adsabs.harvard.edu/abs/2022PhRvL.128y5003B}
}

@ARTICLE{Hakobyan2025,
       author = {{Hakobyan}, Hayk and {B{\"o}ss}, Ludwig M. and {Cai}, Yangyang and {Chernoglazov}, Alexander and {Galishnikova}, Alisa and {Gorbunov}, Evgeny A. and {Mahlmann}, Jens F. and {Philippov}, Alexander and {Solanki}, Siddhant and {Vanthieghem}, Arno and {Zhou}, Muni},
        title = "{Entity -- Hardware-agnostic Particle-in-Cell Code for Plasma Astrophysics. I: Curvilinear Special Relativistic Module}",
      journal = {arXiv e-prints},
         year = 2025,
        month = nov,
          eid = {arXiv:2511.17710},
        pages = {arXiv:2511.17710},
          doi = {10.48550/arXiv.2511.17710},
archivePrefix = {arXiv},
       eprint = {2511.17710},
 primaryClass = {astro-ph.HE},
       adsurl = {https://ui.adsabs.harvard.edu/abs/2025arXiv251117710H}
}

@ARTICLE{Bochenek2020,
       author = {{Bochenek}, C.~D. and {Ravi}, V. and {Belov}, K.~V. and {Hallinan}, G. and {Kocz}, J. and {Kulkarni}, S.~R. and {McKenna}, D.~L.},
        title = "{A fast radio burst associated with a Galactic magnetar}",
      journal = {\nat},
         year = 2020,
        month = nov,
       volume = {587},
       number = {7832},
        pages = {59-62},
          doi = {10.1038/s41586-020-2872-x},
archivePrefix = {arXiv},
       eprint = {2005.10828},
 primaryClass = {astro-ph.HE},
       adsurl = {https://ui.adsabs.harvard.edu/abs/2020Natur.587...59B}
}

@ARTICLE{Bernardi2025,
       author = {{Bernardi}, Dominic and {Yuan}, Yajie and {Chen}, Alexander Y.},
        title = "{Global Kinetic Simulations of Monster Shocks and Their Emission}",
      journal = {\prl},
         year = 2025,
        month = dec,
       volume = {135},
       number = {26},
          eid = {265201},
        pages = {265201},
          doi = {10.1103/y9p7-1zms},
archivePrefix = {arXiv},
       eprint = {2506.04175},
 primaryClass = {astro-ph.HE},
       adsurl = {https://ui.adsabs.harvard.edu/abs/2025PhRvL.135z5201B}
}

@ARTICLE{Grehan2026,
       author = {{Grehan}, Michael P. and {Ripperda}, Bart and {Beloborodov}, Andrei M. and {Thompson}, Christopher and {Most}, Elias R.},
        title = "{Global Magnetohydrodynamic Simulations of Monster Shocks in Neutron Star Magnetospheres}",
      journal = {\apj},
         year = 2026,
        month = jul,
       volume = {1005},
       number = {2},
          eid = {158},
        pages = {158},
          doi = {10.3847/1538-4357/ae633a},
archivePrefix = {arXiv},
       eprint = {2602.21290},
 primaryClass = {astro-ph.HE},
       adsurl = {https://ui.adsabs.harvard.edu/abs/2026ApJ..1005..158G}
}

@ARTICLE{mahlmann2026,
       author = {{Mahlmann}, Jens F. and {Philippov}, Alexander A. and {Soudais}, Adrien and {Beloborodov}, Andrei M. and {Sironi}, Lorenzo},
        title = "{1D PIC Simulations of Resonant Scattering-Driven Pair Cascades in Magnetar Magnetospheres}",
      journal = {arXiv e-prints},
         year = 2026,
        month = sep,
          eid = {arXiv:2609.17661},
        pages = {arXiv:2609.17661},
          doi = {10.48550/arXiv.2609.17661},
archivePrefix = {arXiv},
       eprint = {2609.17661},
 primaryClass = {astro-ph.HE},
       adsurl = {https://ui.adsabs.harvard.edu/abs/2026arXiv260917661M}
}

@ARTICLE{solanki2026,
       author = {{Solanki}, Siddhant and {Mahlmann}, Jens and {Philippov}, Alexander and {Beloborodov}, Andrei},
        title = "{Damping of Fast Radio Bursts in the Inner Magnetospheres of Magnetars}",
      journal = {arXiv e-prints},
         year = 2026,
        month = jun,
          eid = {arXiv:2606.19448},
        pages = {arXiv:2606.19448},
          doi = {10.48550/arXiv.2606.19448},
archivePrefix = {arXiv},
       eprint = {2606.19448},
 primaryClass = {astro-ph.HE},
       adsurl = {https://ui.adsabs.harvard.edu/abs/2026arXiv260619448S}
}

@ARTICLE{golbraikh2023,
       author = {{Golbraikh}, Ephim and {Lyubarsky}, Yuri},
        title = "{On the Escape of Low-frequency Waves from Magnetospheres of Neutron Stars}",
      journal = {\apj},
         year = 2023,
        month = nov,
       volume = {957},
       number = {2},
          eid = {102},
        pages = {102},
          doi = {10.3847/1538-4357/acfa78},
archivePrefix = {arXiv},
       eprint = {2309.09218},
 primaryClass = {astro-ph.HE},
       adsurl = {https://ui.adsabs.harvard.edu/abs/2023ApJ...957..102G}
}

@ARTICLE{sobacchi2025,
       author = {{Sobacchi}, Emanuele},
        title = "{Absorption of strong electromagnetic waves in magnetized pair plasmas}",
      journal = {\pre},
         year = 2025,
        month = dec,
       volume = {112},
       number = {6},
          eid = {065208},
        pages = {065208},
          doi = {10.1103/1b36-qn66},
archivePrefix = {arXiv},
       eprint = {2512.05281},
 primaryClass = {astro-ph.HE},
       adsurl = {https://ui.adsabs.harvard.edu/abs/2025PhRvE.112f5208S}
}

\section*{Appendices}
\appendix

\section{Scaling of the soliton compression}\label{app:scaling}

\begin{figure*}
	\begin{center}
		\includegraphics[width=1\textwidth]{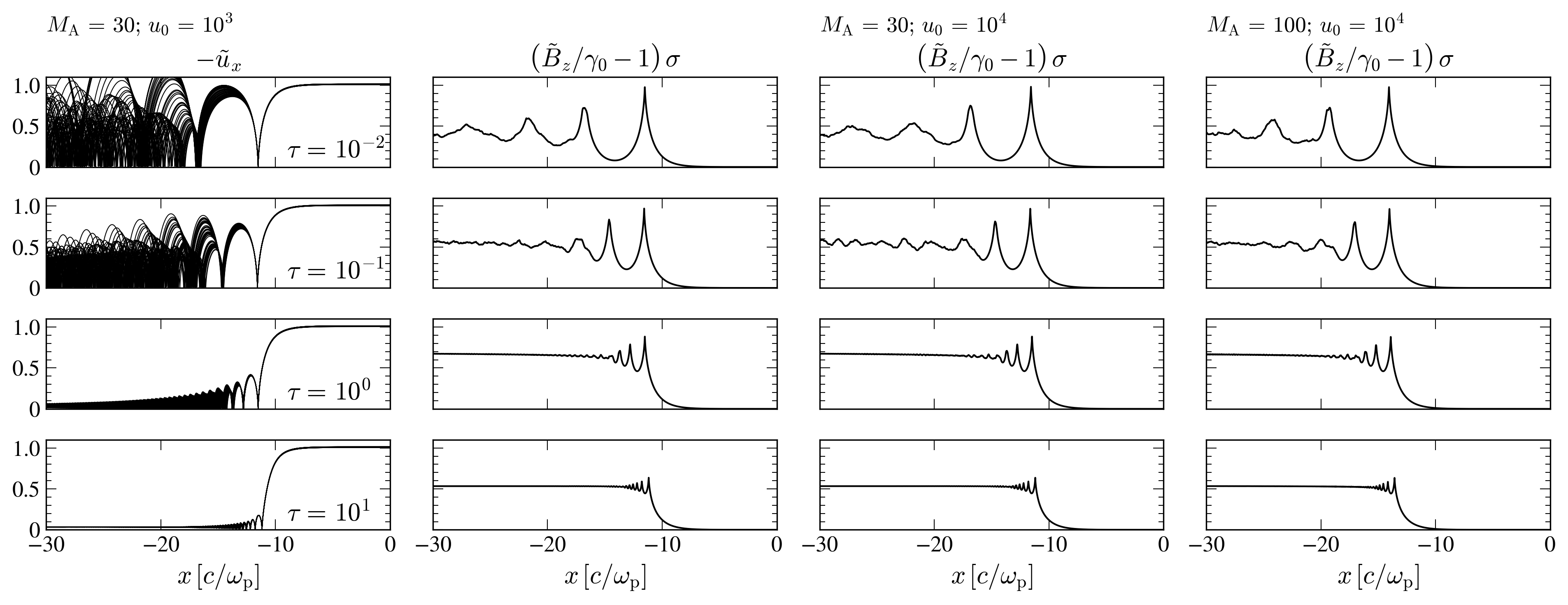}
		\caption{Collapse of the multifluid shock profiles for $\tau = 10^{-2}$, $10^{-1}$, $1$ and $10$ (top to bottom). \emph{Left column:} beam velocities $-\tilde{u}^i_x$ for $(\Ma, u_0) = (30, 10^3)$. \emph{Other columns:} magnetic compression $(\tilde{B}_z/\gamma_0 - 1)\,\sigma_{\rm u}$ for $(\Ma, u_0) = (30, 10^3)$, $(30, 10^4)$ and $(100, 10^4)$, corresponding to $\sigma_{\rm u} \simeq 1.1\times10^3$, $1.1\times10^5$ and $10^4$. At fixed $\tau$, the profiles are nearly identical, confirming that the compression scales as $\sigma_{\rm u}^{-1}$.}
		\label{fig:scale_2}
	\end{center}
\end{figure*}

In Sec.~\ref{sec:theory} we showed that, for $\Ma \gg 1$, the peak compression of the soliton scales as $\delta B_z^{\rm max}/B_0 \simeq \alpha(\tau)/\sigma_{\rm u}$, where $\alpha$ depends on $\tau$ only. Figure~\ref{fig:scale_2} tests this prediction for $\tau = 10^{-2}$--$10$ and three combinations, $(u_0,\Ma) = (10^3, 30)$, $(10^4, 30)$ and $(10^4, 100)$, which span two decades in magnetization, $\sigma_{\rm u} \simeq 1.1\times10^3$--$1.1\times10^5$. All cases lie above the critical Mach number for $\tau \leq 1$, whereas for $\tau = 10$ the cases with $\Ma = 30$ lie just below $\Ma^{\rm crit} \simeq 37.5$, Eq.~\eqref{eq:M_A_crit}. For each value of $\tau$, the three profiles are nearly indistinguishable: both the peak compression at the shock transition and the envelope of the downstream soliton train collapse onto a single curve.

The left column of Fig.~\ref{fig:scale_2} shows the corresponding beam velocities for $(u_0,\Ma) = (10^3, 30)$. For $\tau \lesssim 10^{-1}$, the beams phase-mix within a few soliton crossings, as in the uncooled case \citep{vanthieghem_relativistically_2025}. For $\tau = 1$, the solitary structure is progressively damped while phase mixing still leads to the formation of a shrunk ring distribution. For $\tau = 10$, the beams remain laminar and follow a single damped trajectory, as expected below the critical Mach number.

\section{Comparison with kinetic simulations}\label{app:pic}

\begin{figure*}
    \centering
    \begin{minipage}[c]{0.4\textwidth}
        \centering
        \includegraphics[width=\textwidth]{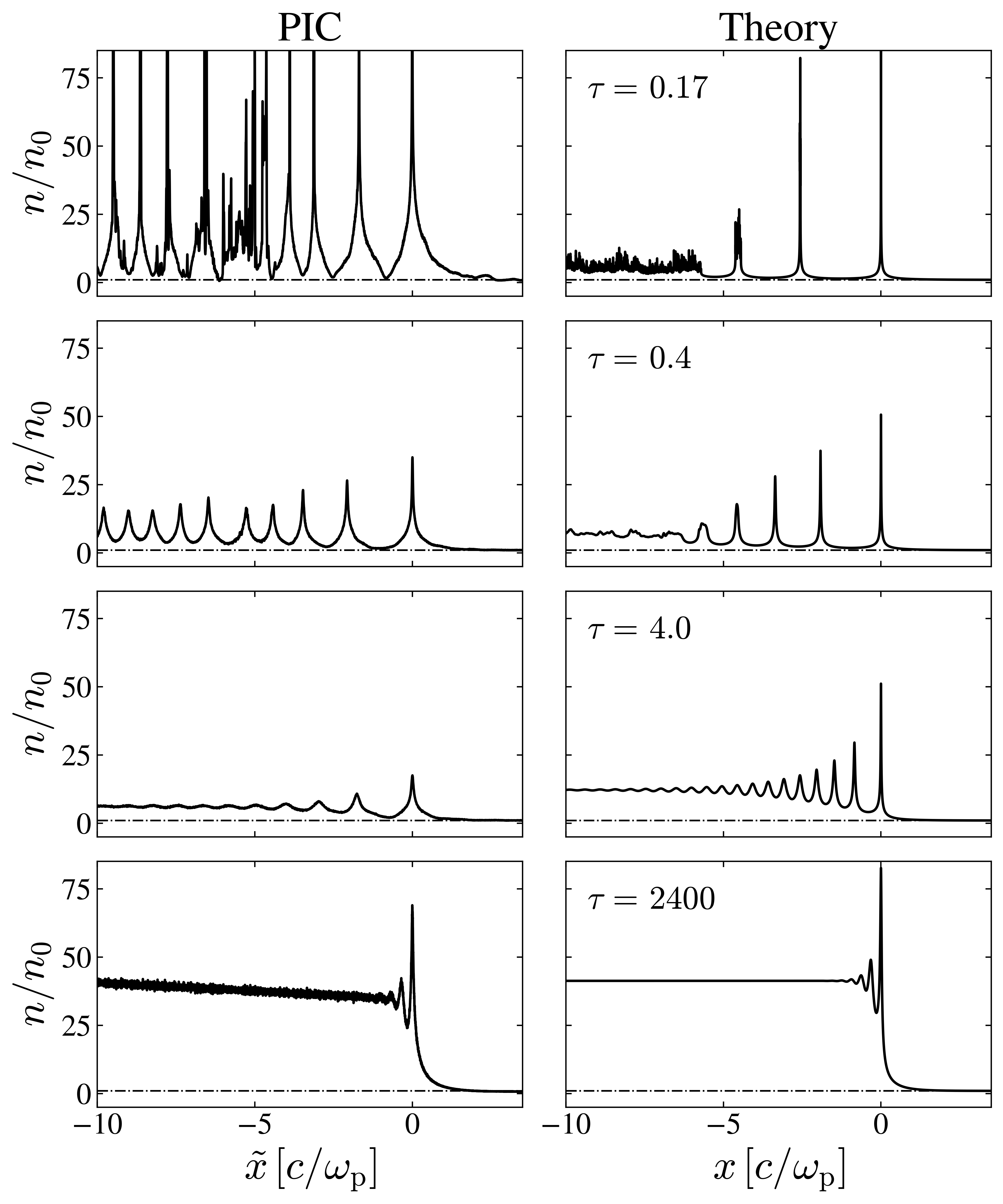}
    \end{minipage}
    \hfill
    \begin{minipage}[c]{0.52\textwidth}
        \caption{
        Proper density profiles, normalized to their upstream value, from PIC simulations $\mathcal{S}_2$--$\mathcal{S}_5$ at $\wpp t = 9750$ (left), compared with multifluid solutions with $N = 100$ beams computed for the upstream parameters measured in each simulation (right). The effective values of the cooling parameter are $\tau \simeq 0.17$, $0.4$, $4.0$ and $2.4\times10^3$, from top to bottom. The PIC abscissa is rescaled as $\tilde{x} = \gamma_{\rm sh|\star}\,x$, with $\gamma_{\rm sh|\star} \simeq 8$ being the shock Lorentz factor in the simulation frame, to account for relativistic length contraction.
        }
        \label{fig:comp}
    \end{minipage}
\end{figure*}

Figure~\ref{fig:comp} compares the proper density profiles of simulations $\mathcal{S}_2$--$\mathcal{S}_5$ at $\wpp t = 9750$ with multifluid solutions ($N = 100$) computed for the upstream parameters measured in each simulation immediately ahead of the shock. The effective cooling parameters at this time are $\tau \simeq 0.17$, $0.4$, $4.0$ and $2.4\times10^3$, respectively. The PIC abscissa is rescaled as $\tilde{x} = \gamma_{\rm sh|\star}\,x$, with $\gamma_{\rm sh|\star} \simeq 8$ the shock Lorentz factor in the simulation frame, to account for the length contraction between the simulation and shock frames.

The stationary model captures the main features of the kinetic shocks. For $\mathcal{S}_3$, it reproduces the positions of the leading solitons and the progressive shortening of their spacing. For $\mathcal{S}_5$, where the soliton train is quenched, it reproduces the sharp density spike at the shock front and the downstream plateau, $n/n_0 \simeq 40$. The main discrepancies occur for $\mathcal{S}_2$ and $\mathcal{S}_4$. In $\mathcal{S}_2$, the kinetic profile exhibits a longer and more intense soliton train than the stationary solution, whose beams phase-mix after about three solitons. In $\mathcal{S}_4$, the stationary solution predicts a more closely spaced train, and overestimates the soliton amplitudes. These differences may reflect effects associated with the proper boost of the 4-current and backreaction of the coherent emission on the shock structure.

\section{Field-derivative terms of the Landau--Lifshitz force}\label{app:LL}

\begin{figure*}
	\begin{center}
		\includegraphics[width=1\textwidth]{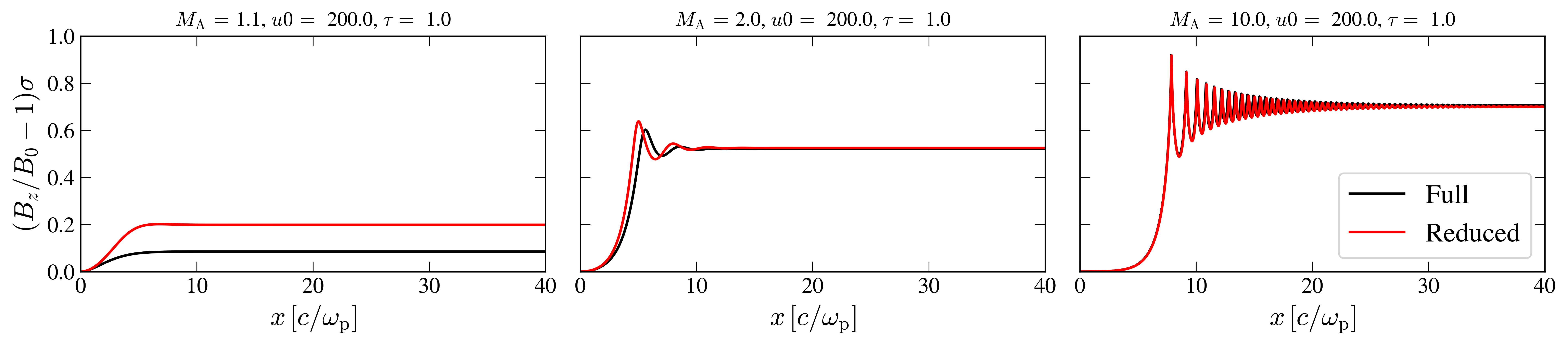}
		\caption{
        Magnetic compression $(B_z/B_0 - 1)\,\sigma_{\rm u}$ obtained with the full (black) and reduced (red) Landau--Lifshitz force, for $\Ma = 1.1$, $2$ and $10$ at fixed $u_0 = 200$ and $\tau = 1$.}
		\label{fig:scale_llr}
	\end{center}
\end{figure*}

The radiation-reaction force of Eq.~\eqref{eq:RRforce}, which is also the one implemented in our PIC simulations, is the reduced form of the Landau--Lifshitz force, in which the terms involving derivatives of the electromagnetic field are neglected. Including this term in the PIC simulations is beyond the scope of this work, but its effect on the shock structure can be assessed with the stationary multifluid model. In the geometry of Sec.~\ref{sec:theory}, $\partial_t = 0$ and the electric field is uniform, so that only the convective derivative of the magnetic field contributes, $\bm{\mathcal{F}}^i_\partial = (\sigma_{\rm T}/4\pi e)\, u^i_x\, \partial_x B_z \,(\beta^i_y\,\hat{x} - \beta^i_x\,\hat{y})$. Dividing by $\beta^i_x$ and adopting the normalization of Sec.~\ref{sec:theory}, the right-hand sides of Eqs.~\eqref{eq:1_fluid_1} and \eqref{eq:1_fluid_2} acquire the additional terms
\begin{align}
    \Delta^i_x \,&=\, \frac{\tau}{\Ma^3}\, \tilde{u}^i_y\, \partial_{\tilde{x}} \tilde{B}_z \,, \label{eq:Dx}\\
    \Delta^i_y \,&=\, -\frac{\tau}{\Ma^3}\, \tilde{u}^i_x\, \partial_{\tilde{x}} \tilde{B}_z \, \label{eq:Dy} \,.
\end{align}
The effect on the scaling of the maximum compression of the soliton is shown in Fig.~\ref{fig:scale_llr}. The profile only deviates for large $\tau$ and low $\Ma$. 

\section{Coupling between the precursor wave and the upstream flow}\label{app:coupling}

When cooling is strong but the shock remains above the emission threshold, $\Ma > \Ma^{\rm crit}$, the precursor wave is emitted and can decelerate the incoming plasma, so that $\gamma_{\rm u}$ falls below the value given by Eq.~\eqref{eq:scal1} and the upstream magnetization increases by the same factor. The scaling relations of Sec.~\ref{sec:monster} should therefore be regarded as upper bounds. In monster shocks, the precursor wave is in the strongly nonlinear regime, $a_0 \gg 1$ (Eq.~\eqref{eq:ao}), and its interaction with the plasma ahead of the shock could decelerate the upstream flow and self-regulate the emission~\citep{Beloborodov_2026}. Interestingly, our simulations show a similar deceleration, although the precursor wave remains in the linear regime, $a_0 \ll 1$. The upstream Lorentz factor of $\mathcal{S}_3$ and $\mathcal{S}_4$ remains below that of both the weakly cooled run $\mathcal{S}_1$ and the strongly cooled run $\mathcal{S}_5$ (Figs.~\ref{fig:MS_global} and~\ref{fig:MS_prof}). In this regime, the deceleration results from the radiative damping of the precursor wave, which transfers its momentum to the incoming flow as it propagates into the upstream plasma. Linearizing the two-fluid equations with the radiation-reaction force of Eq.~\eqref{eq:RRforce} about a cold magnetized pair plasma, we obtain the dispersion relation of extraordinary modes in the upstream rest frame:
\begin{align} \label{eq:disp_rr}
\frac{\tilde{k}^2}{\tilde{\omega}^2} \,=\, 1\,-\,\frac{1}{\left(\tilde{\omega} + i\,\tilde{\nu}\right)^2 - \sigma_{\rm u}}\,,
\end{align}
where $\tilde{\omega} = \omega'/\omega'_{\rm p}$ and $\tilde{k} = k' c/\omega'_{\rm p}$ are normalized to the proper plasma frequency. Radiation reaction thus enters as an effective collision frequency, $\tilde{\nu} = \nu/\omega'_{\rm p} = \sigma_{\rm u}\,\sigma_{\rm T} n_{\rm u} c/\omega'_{\rm p}$, where $\nu$ is the radiative damping rate of the particle gyration. We stress that, at linear order, the damping originates entirely from the $\bm{\kappa}$ term of Eq.~\eqref{eq:RRforce}, the drag term being of second order in the wave amplitude. For $\tilde{\omega} \gg \sqrt{\sigma_{\rm u}}$, the wave amplitude decays over a length $\ell_{\rm abs} \simeq (\tilde{\omega}^2/\tilde{\nu})\,c/\omega'_{\rm p}$ in the upstream rest frame. The deceleration is therefore most pronounced at intermediate cooling, where the precursor wave is still emitted but efficiently absorbed ahead of the shock, consistent with the reduction of $\gamma_{\rm u}$ in $\mathcal{S}_3$ and $\mathcal{S}_4$.

Although the coupling in monster shocks is nonlinear, its impact on the shock parameters can be estimated without specifying the underlying mechanism. We define a deceleration factor $\eta = \hat{\gamma}_{\rm u}/\gamma_{\rm u} < 1$, where hatted quantities refer to the shock modified by the precursor wave, and assume that the lab-frame magnetic field and density ahead of the shock are unaffected, so that $\hat{\sigma}_{\rm u} = \sigma_{\rm u}/\eta$ and $\hat{n}_{\rm u} = n_{\rm u}/\eta$. At fixed radius and for $\hat{\tau} \gg 1$, the scaling relations of Eq.~\eqref{eq:scal1} then become
\begin{align} \label{eq:eta_scal}
    \hat{\tau} \,&\simeq\, \eta^{2}\,\tau\,, \\
    \hat{M}_{\rm A} \,&\simeq\,  \eta\,\Ma\,,\\
    \frac{\hat{M}_{\rm A}}{\hat{M}_{\rm A}^{\rm crit}} \,&\simeq\, \eta^{1/10}\,\frac{\Ma}{\Ma^{\rm crit}}\,, \\
    \hat{\nu}_{\rm FRB} \,&\simeq\,  \eta^{-3/7}\,\nu_{\rm FRB}\,.
\end{align}
The deceleration therefore weakens the cooling, but hardly affects the ratio $\Ma/\Ma^{\rm crit}$. For $\eta = 0.1$, the cooling parameter drops by two orders of magnitude, whereas $\Ma/\Ma^{\rm crit}$ decreases by only $\approx 20\%$ and remains of order $10^3$, so that our conclusion that radiation reaction does not quench the precursor wave is robust to the coupling with the upstream flow. In contrast, the peak frequency increases as $\eta^{-3/7}$, by a factor of about $3$ for $\eta = 0.1$, which pushes the emission further into the kinetic regime. Note, however, that the ratio $\gamma_{\rm u}/a_0$ of Eq.~\eqref{eq:gamma/a} also decreases, as $\eta^{4/7}$ once the upshift of the precursor frequency is included, which brings the onset of synchrotron absorption, $\omega' \sim a_0\,\omega'_B$, closer to the dissipation zone. A self-consistent determination of $\eta$ requires a nonlinear treatment of the coupling, which is left for future work.

\end{document}